\documentclass[prd,aps,nofootinbib,preprintnumbers,amsmath,amssymb,floatfix]{revtex4}

\usepackage[a0paper]{geometry}
\usepackage{anysize}
\marginsize{2.5cm}{2cm}{2.5cm}{2.5 cm}
\usepackage{amsmath}
\usepackage{amsfonts}
\usepackage{amssymb}
\usepackage{slashed}
\usepackage{mathrsfs}
\usepackage{epsfig}
\usepackage{enumerate}
\usepackage{amsmath}
\usepackage{pdfpages}
\usepackage{setspace}
\usepackage[utf8]{inputenc}
\usepackage{float}
\usepackage{ytableau}
\usepackage{color}
\usepackage{times}
\usepackage{hyperref}

\usepackage{graphicx}
\usepackage{dcolumn}
\usepackage{bm}
\usepackage{slashed}

\newcommand{\pr}[1]{\mathcal{P}^{(#1)}}

\begin{document}

\title{An alternative approach to baryon masses in the $1/N_c$ expansion of QCD}

\author{
Rub\'en Flores-Mendieta
}
\affiliation{
Instituto de F{\'\i}sica, Universidad Aut\'onoma de San Luis Potos{\'\i}, \'Alvaro Obreg\'on 64, Zona Centro, San Luis Potos{\'\i}, S.L.P.\ 78000, Mexico
}

\author{
Sergio Alejandro Garc{\'\i}a-Monreal
}
\affiliation{
Instituto de F{\'\i}sica, Universidad Aut\'onoma de San Luis Potos{\'\i}, \'Alvaro Obreg\'on 64, Zona Centro, San Luis Potos{\'\i}, S.L.P.\ 78000, Mexico
}

\author{
Luis Rom\'an Ruiz-Robles
}
\affiliation{
Instituto de F{\'\i}sica, Universidad Aut\'onoma de San Luis Potos{\'\i}, \'Alvaro Obreg\'on 64, Zona Centro, San Luis Potos{\'\i}, S.L.P.\ 78000, Mexico
}

\author{
Francisco Alberto Torres-Bautista
}
\affiliation{
Instituto de F{\'\i}sica, Universidad Aut\'onoma de San Luis Potos{\'\i}, \'Alvaro Obreg\'on 64, Zona Centro, San Luis Potos{\'\i}, S.L.P.\ 78000, Mexico
}

\date{\today}

\begin{abstract}
The baryon mass operator is studied within a combined expansion in $1/N_c$ and perturbative $SU(3)$ flavor symmetry breaking, where $N_c$ denotes the number of quark charges. Flavor projection operators are used to classify the baryon operators involved in the expansion, which fall into the flavor representations $1$, $8$, $10+\overline{10}$, $27$, $35+\overline{35}$ and $64$. This approach allows one to incorporate up to third-order flavor symmetry breaking in the baryon mass operator in a rigorous and systematic way. Previous work on the subject is considered to validate the approach. A fit to data is performed to evaluate the free parameters in the theory and to produce some numerical values of baryon masses. Results are consistent and reaffirm the striking success of the $1/N_c$ expansion.
\end{abstract}

\maketitle

\section{Introduction}

The $SU(3)$ flavor symmetry of the strong interaction suggested by Gell-Mann \cite{gell-mann} and Ne’eman \cite{yn} in the early 60’s of the past century has undoubtedly become the most successful organizational scheme for hadrons to the extent that it played a crucial role in the development of the quark model. Hadrons were thus organized into $SU(3)$ representation multiplets---octets and decuplets. $SU(3)$ flavor symmetry is, of course, an approximate one: It is broken in QCD by non-equal masses of the up and down quarks and the strange quark. It is thus postulated that the $SU(3)$ violating part of the Hamiltonian transforms like the eighth component of an adjoint (octet) representation of $SU(3)$ with zero isospin and hypercharge. Important consequences of $SU(3)$ flavor symmetry breaking (SB) can be seen in the Gell-Mann–Okubo (GMO) mass formulas describing the mass splitting inside a given $SU(3)$ multiplet.

In the early studies of nuclear reactions it was observed that, to a good approximation, the strong interaction is independent of the electric charge carried by nucleons, so it is invariant under a transformation which interchanges proton and neutron. In modern terminology, isospin is regarded as a symmetry of the strong interaction under the action of the group $SU(2)$, the two states being the up and down quarks, with $m_u=m_d$. Since different members of a given isospin multiplet have different electric charges, the electromagnetic interaction clearly does not respect the isospin symmetry. Thus, isospin symmetry breaking originates from two different sources: Electromagnetic self-energies and the difference of the up and down quark masses. The latter is referred to as
strong isospin-breaking and is regarded as the leading contribution.

Nowadays, when comparing theoretical predictions with experimental measurements isospin breaking corrections can not be in general neglected. In particular, isospin symmetry breaking in mass splittings of the lowest-lying (octet and decuplet) baryons is an important issue to be accounted for. Lots of efforts and a considerable number of methods have been devoted to study it from both the analytical and numerical bent. A selection of such methods is constituted by the $1/N_c$ expansion \cite{jen93,jl}, chiral perturbation theory \cite{cpt1,cpt2,cpt3}, a combined expansion in chiral symmetry breaking and $1/N_c$ \cite{jen96}, QCD sum rules \cite{sr1,sr2}, chiral soliton model \cite{sm1}, the fast-growing lattice QCD \cite{romiti,lat1,lat2,lat3,lat4,lat5}, to name but a few.

The present work is devoted to evaluate SB effects in the baryon mass sector of the lowest-lying baryons in the context of the $1/N_c$ expansion of QCD. This subject has already been dealt with in a detail-oriented paper by Jenkins and Lebed \cite{jl}. In that work, special emphasis on the $I=0$, $1$, $2$ and $3$ mass splitting of the octet and decuplet baryons was put in a detailed computation in the $1/N_c$ expansion combined with perturbative SB. A great deal of evidence for the mass hierarchy was found in this combined expansion. Here, an alternative pragmatic strategy to analyze baryon masses within the $1/N_c$ expansion is implemented. In this approach, $SU(3)$ flavor projection operators \cite{banda1,banda2} are widely used. On a first stage, the most general baryon operator basis containing up to three flavor adjoint indices is constructed. Successive contractions of these operators with $SU(3)$ tensors provide the corresponding bases with lesser flavor indices. On a second stage, once all participating operators are identified, projection operators are applied to them in order to get all possible flavor representations that enter into play in SB. Although the approach notably complicates because a huge number of free operator coefficients appear, a thorough analysis allows one to redefine practically all of them in terms of only 21 independent effective free parameters, in total agreement with the analysis of Ref.~\cite{jl}.\footnote{In fact, two additional parameters are identified here which are not apparent in Ref.~\cite{jl}. These new parameters come along with operators containing three adjoint indices in the $10+\overline{10}$ representation and only affect the off-diagonal mass $\Sigma^0\Lambda$.}

The paper is organized as follows. In Sec.~\ref{sec:massop} some basics definitions on the large-$N_c$ limit of QCD are provided in order to set notation and conventions. The baryon mass operator is outlined, first in the $SU(3)$ symmetry limit and then, including SB perturbatively. In Sec.~\ref{sec:opbasis}, the $1/N_c$ expansion is constructed starting from the determination of the operator bases for three, two, one and zero free flavor indices, which are necessary to evaluate up to third-order SB. Flavor projection operators defined in Refs.~\cite{banda1,banda2} are widely used to obtain all the flavor representation allowed in the tensor product of two and three adjoint representations. The resultant irreducible representations are $1$, $8$, $10+\overline{10}$, $27$, $35+\overline{35}$ and $64$. With the matrix elements of the participating operators available, operator coefficients are reorganized to be absorbed into 21 effective operator coefficients. This yields the final expressions for baryon masses in the $1/N_c$ expansion combined with perturbative SB. In Sec.~\ref{sec:num} a least-squares fit to data is performed to explore the 19 relevant free parameters in the analysis, using experimental \cite{part} and numerical \cite{romiti} data. In Sec.~\ref{sec:massrel}, some interesting mass relations falling in the $I=0$, $1$, $2$ and $3$ channels obtained in Ref.~\cite{jl} are tested. In Sec.~\ref{sec:con} some concluding remarks are given. The paper is complemented by two appendices. In Appendix \ref{app:q1q2q3}, explicit expressions for the flavor projection operators acting on the product of three adjoints are presented. In Appendix \ref{app:fullmass}, explicit expressions for the baryon masses in terms of the operator coefficients are listed. These expressions are the ones that can be used in actual least-squares fits to data.

\section{\label{sec:massop}Baryon mass operator in large-$N_c$ QCD}

In this section, a few facts on the large-$N_c$ limit of QCD are given in order to set notation and conventions. Further technical aspects can be found in the original papers \cite{tHooft,ven,witten,dm1,djm94,djm95} and references therein.

The $1/N_c$ expansion of any baryon operator transforming according to a given $SU(2)\times SU(3)$ representation can be written as \cite{djm95}
\begin{equation}
\mathcal{O} = \sum_n c_n \frac{1}{N_c^{n-1}} \mathcal{O}_n, \label{eq:1ncE}
\end{equation}
where the $\mathcal{O}_n$ constitute a complete set of linearly independent effective $n$-body operators which can be written as polynomials in the $SU(6)$ generators
\begin{equation}
J^k = q^\dagger \frac{\sigma^k}{2} q, \qquad T^c = q^\dagger \frac{\lambda^c}{2} q, \qquad G^{kc} = q^\dagger
\frac{\sigma^k}{2}\frac{\lambda^c}{2} q, \label{eq:su6gen}
\end{equation}
where $J^k$ are spin generators, $T^c$ are flavor generators, and $G^{kc}$ are spin-flavor generators which satisfy the commutation relations listed in Table \ref{tab:surel} \cite{djm95}. Here $q^\dagger$ and $q$ are $SU(6)$ operators that create and annihilate states in the fundamental representation of $SU(6)$, and $\sigma^k$ and $\lambda^c$ are the Pauli spin and Gell-Mann flavor matrices, respectively.
\begingroup
\begin{table}
\caption{\label{tab:surel}$\mathrm{SU}(2 N_f)$ commutation relations.}
\bigskip
\label{tab:su2fcomm}
\centerline{\vbox{ \tabskip=0pt \offinterlineskip
\halign{
\strut\quad $ # $\quad\hfil&\strut\quad $ # $\quad \hfil\cr
\multispan2\hfil $\left[J^i,T^a\right]=0,$ \hfil \cr
\noalign{\medskip}
\left[J^i,J^j\right]=i\epsilon^{ijk} J^k,
&\left[T^a,T^b\right]=i f^{abc} T^c,\cr
\noalign{\medskip}
\left[J^i,G^{ja}\right]=i\epsilon^{ijk} G^{ka},
&\left[T^a,G^{ib}\right]=i f^{abc} G^{ic},\cr
\noalign{\medskip}
\multispan2\hfil$\displaystyle [G^{ia},G^{jb}] = \frac{i}{4}\delta^{ij}
f^{abc} T^c + \frac{i}{2N_f} \delta^{ab} \epsilon^{ijk} J^k + \frac{i}{2} \epsilon^{ijk} d^{abc} G^{kc}.$ \hfill\cr
}}}
\end{table}
\endgroup

In the large-$N_c$ limit of QCD, one of the earliest analyses of the masses of the $J^P = \frac12^+$ and $J^P = \frac32^+$ physical baryons ---hereafter referred to as octet and decuplet baryons, respectively--- proved them to be proportional to $J^2$ \cite{jen93}. Later work, dealing with the $I = 0,1,2,3$ baryon mass splittings in a systematic expansion in $1/N_c$ and perturbative $SU(3)$ flavor symmetry breaking, found a remarkable evidence for the observed mass hierarchy \cite{jl}.

The baryon mass operator, hereafter denoted by $M$, transforms as $(0,1)$ under the $SU(2)\times SU(3)$ spin-flavor symmetry. In the flavor symmetry limit, $M$ is given by \cite{djm95}
\begin{equation}
M = N_c \mathcal{P}\left( \frac{J^2}{N_c^2} \right),
\end{equation}
where $\mathcal{P}$ stands for a polynomial. For $N_c = 3$, the specific form of the $1/N_c$ expansion of $M$ reads,
\begin{equation}
M = m_0 N_c \openone + \frac{1}{N_c} m_2 J^2, \label{eq:su3mass}
\end{equation}
where $m_0$ and $m_2$ are unknown parameters. Thus, the baryon mass is of order $\mathcal{O}(N_c)$, since it contains $N_c$ quarks.

\subsection{Baryon mass operator including perturbative SB}

Due to the different light quark masses, $SU(3)$ flavor symmetry is broken. There are two sources of SB. The first one is due to the light quark masses and the perturbation transforms as the adjoint (octet) irreducible representation of $SU(3)$,
\begin{equation}
\epsilon \mathcal{H}^8 + \epsilon^\prime \mathcal{H}^3. \label{eq:sb}
\end{equation}
The first summand in Eq.~(\ref{eq:sb}) represents the dominant $SU(3)$ breaking and transforms as the eighth component of a flavor octet,
where $\epsilon \sim m_s/{\Lambda_{\mathrm{QCD}}}$ is a (dimensionless) measure of SB. The second summand represents the leading QCD isospin breaking effect, i.e., the one associated with the difference of the up and down quark masses and transforms as the third component of a flavor octet, where $\epsilon^\prime \sim (m_d-m_u)/\Lambda_{\mathrm{QCD}}$, so $\epsilon^\prime \ll \epsilon$. The effects of SB in the baryon masses within the $1/N_c$ expansion have been meticulously discussed in Ref.~\cite{jl}, where it was pointed out that while the baryon mass is about 1 GeV, isospin mass splittings typically round several MeV, so $\epsilon^\prime$ represents breaking effects of order $1/N_c^5$ in QCD. The second source of SB is induced by electromagnetic interactions. Electromagnetic mass splittings are second order in the quark charge matrix so they get a suppression factor of $\epsilon^{\prime\prime} \sim \alpha_{\mathrm{em}}/4\pi$ \cite{jl}. These splittings round a few MeV so to a good approximation
\begin{equation}
\frac{m_d - m_u}{\Lambda_{\mathrm{QCD}}} \sim \frac{\alpha_{\mathrm{em}}}{4\pi}.
\end{equation}

The starting point of the analysis of Ref.~\cite{jl} was the construction of the relevant $1/N_c$ expansions, classified into isospin channels $I=0$, $1$, $2$ and $3$. At first order in $SU(3)$ breaking, the baryon mass term transforms as an $SU(3)$ octet. The most general spin-zero $SU(3)$ octet is a polynomial in $J^i$, $T^a$ and $G^{ia}$, with one free flavor index set to either 3 or 8. Thus, $O^3$ and $O^8$ operators correspond to $I=0$ and $I=1$, respectively. At second order in $SU(3)$ breaking, a tensor with two free flavor indices should be obtained. Relevant operators for baryon mass splittings are $O^{88}$, $O^{38}$ and $O^{33}$ with $I=0$, $1$ and $2$, respectively. Similarly, at third order, a tensor with three free flavor indices should be obtained so the relevant operators are in this case $O^{888}$, $O^{388}$, $O^{338}$ and $O^{333}$ with $I=0$, $1$, $2$ and $3$, respectively. Electromagnetic corrections only appear in the $I=0$, $1$, and $2$ channels, so contributions of the form $O^{888}$ amount corrections of order $\epsilon^3$ alone.

The construction of the $1/N_c$ expansions for all isospin channels are provided in the following section, using $SU(3)$ flavor projection operators as introduced in Refs.~\cite{banda1,banda2} as an alternative approach to the problem.

\section{\label{sec:opbasis}Construction of the $1/N_c$ expansion for the baryon mass operator}

From a group theory point of view, symmetry breaking can be incorporated in the analysis of a baryon operator ({\it v.gr}. mass, axial and vector current, magnetic moment, etc.) by considering multiple tensor products of $SU(3)$ flavor octets. At first order in $SU(3)$ breaking, the baryon mass term transforms as an $SU(3)$ octet. At second and third order SB, it is found that \cite{banda2}
\begin{equation}
\mathbf{8} \otimes \mathbf{8} = \mathbf{1} \oplus 2(\mathbf{8}) \oplus \mathbf{10} \oplus \overline{\mathbf{10}} \oplus \mathbf{27}, \label{eq:8x8}
\end{equation}
and
\begin{equation}
\mathbf{8} \otimes \mathbf{8} \otimes \mathbf{8} = 2(\mathbf{1}) \oplus 8(\mathbf{8}) \oplus 4(\mathbf{10} \oplus \overline{\mathbf{10}}) \oplus 6(\mathbf{27}) \oplus 2(\mathbf{35} \oplus \overline{\mathbf{35}}) \oplus \mathbf{64}, \label{eq:8x8x8}
\end{equation}
respectively, where the right-hand sides denote the dimensions of the irreducible representations of $SU(3)$.

The analysis of the baryon mass splittings in the physical baryons of Ref.~\cite{jl} thus provided the $1/N_c$ expansions for the $SU(2) \times SU(3)$ representations $(0,1)$, $(0,8)$, $(0,10+\overline{10})$, $(0,27)$, and $(0,64)$, since the baryon $1/N_c$ expansion extends only to $3$-body operators restricting the analysis to the physical baryons.

In this section the baryon mass splittings of the physical baryons are analyzed following a more pragmatic approach. On a first stage, the most complete operator basis containing spin-0 objects with up to $3$-body operators with three flavor indices are constructed. Proper contractions of flavor indices are thus performed to obtain the corresponding bases with two, one, and zero flavor indices. On a second stage, the use of $SU(3)$ flavor projection operators introduced in Refs.~\cite{banda1,banda2} will allow one to identify unambiguously all flavor representations relevant in the analysis of baryon mass splittings.

\subsection{Operator bases}

In order to obtain the $1/N_c$ expansion of the baryon mass operator including SB terms, the operator bases containing spin-0 objects with zero, one, two, and three flavor indices should be constructed. Let ${\sf M}$, ${\sf M}^a$, ${\sf M}^{a_1a_2}$ and ${\sf M}^{a_1a_2a_3}$ denote such operator bases. A previous analysis on baryon-meson scattering \cite{banda2} introduced the operator basis $R^{(ij)(a_1a_2a_3)}$, which is constituted by 170 linearly independent spin-2 objects with three flavor indices, retaining up to 3-body operators. In the present case, the bases are thus obtained by contracting the spin and flavor indices on $R^{(ij)(a_1a_2a_3)}$ using spin and flavor invariant tensors, such as $\delta^{ij}$, $\delta^{a_1a_2}$, $if^{a_1a_2a_3}$, or $d^{a_1a_2a_3}$, as the case may be. Their explicit forms are given in the following sections.

\subsubsection{${\sf M}^{a_1a_2a_3}$ basis}

The ${\sf M}^{a_1a_2a_3}$ basis is obtained from $R^{(ij)(a_1a_2a_3)}$ by simply contracting the spin indices with $\delta^{ij}$. After removing redundant operators, the resultant basis is
\begin{equation}
{\sf M}^{a_1a_2a_3} = \{M_i^{a_1a_2a_3}\}, \label{eq:mabc}
\end{equation}
where
\begin{eqnarray}
\begin{array}{lll}
M_{1}^{a_1a_2a_3} = i f^{a_1a_2a_3}, & \qquad &
M_{2}^{a_1a_2a_3} = d^{a_1a_2a_3}, \\
M_{3}^{a_1a_2a_3} = \delta^{a_1a_2} T^{a_3}, & \qquad &
M_{4}^{a_1a_2a_3} = \delta^{a_1a_3} T^{a_2}, \\
M_{5}^{a_1a_2a_3} = \delta^{a_2a_3} T^{a_1}, & \qquad &
M_{6}^{a_1a_2a_3} = i f^{a_1a_2a_3} J^2, \\
M_{7}^{a_1a_2a_3} = d^{a_1a_2a_3} J^2, & \qquad &
M_{8}^{a_1a_2a_3} = \delta^{a_1a_2} \{J^r,G^{ra_3}\}, \\
M_{9}^{a_1a_2a_3} = \delta^{a_1a_3} \{J^r,G^{ra_2}\}, & \qquad &
M_{10}^{a_1a_2a_3} = \delta^{a_2a_3} \{J^r,G^{ra_1}\}, \\
M_{11}^{a_1a_2a_3} = f^{a_1a_2e_1} f^{a_3e_1g_1} \{J^r,G^{rg_1}\}, & \qquad &
M_{12}^{a_1a_2a_3} = f^{a_1a_3e_1} f^{a_2e_1g_1} \{J^r,G^{rg_1}\}, \\
M_{13}^{a_1a_2a_3} = d^{a_1a_2e_1} d^{a_3e_1g_1} \{J^r,G^{rg_1}\}, & \qquad &
M_{14}^{a_1a_2a_3} = i f^{a_1a_2e_1} d^{a_3e_1g_1} \{J^r,G^{rg_1}\}, \\
M_{15}^{a_1a_2a_3} = i f^{a_1a_3e_1} d^{a_2e_1g_1} \{J^r,G^{rg_1}\}, & \qquad &
M_{16}^{a_1a_2a_3} = i d^{a_1e_1g_1} f^{a_2a_3e_1} \{J^r,G^{rg_1}\}, \\
M_{17}^{a_1a_2a_3} = i f^{a_1a_2e_1} \{T^{a_3},T^{e_1}\}, & \qquad &
M_{18}^{a_1a_2a_3} = d^{a_1a_2e_1} \{T^{a_3},T^{e_1}\}, \\
M_{19}^{a_1a_2a_3} = d^{a_1a_3e_1} \{T^{a_2},T^{e_1}\}, & \qquad &
M_{20}^{a_1a_2a_3} = d^{a_2a_3e_1} \{T^{a_1},T^{e_1}\}, \\
M_{21}^{a_1a_2a_3} = [T^{a_1},\{T^{a_2},T^{a_3}\}], & \qquad &
M_{22}^{a_1a_2a_3} = [T^{a_3},\{T^{a_1},T^{a_2}\}], \\
M_{23}^{a_1a_2a_3} = \delta^{a_1a_2} \{J^2,T^{a_3}\}, & \qquad &
M_{24}^{a_1a_2a_3} = \delta^{a_1a_3} \{J^2,T^{a_2}\}, \\
M_{25}^{a_1a_2a_3} = \delta^{a_2a_3} \{J^2,T^{a_1}\}, & \qquad &
M_{26}^{a_1a_2a_3} = \{T^{a_1},\{T^{a_2},T^{a_3}\}\}, \\
M_{27}^{a_1a_2a_3} = \{T^{a_2},\{T^{a_1},T^{a_3}\}\}, & \qquad &
M_{28}^{a_1a_2a_3} = \{T^{a_3},\{T^{a_1},T^{a_2}\}\}, \\
M_{29}^{a_1a_2a_3} = \{T^{a_1},\{G^{ra_2},G^{ra_3}\}\}, & \qquad &
M_{30}^{a_1a_2a_3} = \{T^{a_2},\{G^{ra_1},G^{ra_3}\}\}, \\
M_{31}^{a_1a_2a_3} = \{T^{a_3},\{G^{ra_1},G^{ra_2}\}\}, & \qquad &
M_{32}^{a_1a_2a_3} = i f^{a_1a_2e_1} \{T^{a_3},\{J^r,G^{re_1}\}\}, \\
M_{33}^{a_1a_2a_3} = i f^{a_1a_3e_1} \{T^{a_2},\{J^r,G^{re_1}\}\}, & \qquad &
M_{34}^{a_1a_2a_3} = i f^{a_2a_3e_1} \{T^{a_1},\{J^r,G^{re_1}\}\}, \\
M_{35}^{a_1a_2a_3} = i f^{a_1a_2e_1} \{T^{e_1},\{J^r,G^{ra_3}\}\}, & \qquad &
M_{36}^{a_1a_2a_3} = i f^{a_1a_3e_1} \{T^{e_1},\{J^r,G^{ra_2}\}\}, \\
M_{37}^{a_1a_2a_3} = i f^{a_2a_3e_1} \{T^{e_1},\{J^r,G^{ra_1}\}\}, & \qquad &
M_{38}^{a_1a_2a_3} = d^{a_1a_2e_1} \{T^{e_1},\{J^r,G^{ra_3}\}\}, \\
M_{39}^{a_1a_2a_3} = d^{a_1a_3e_1} \{T^{e_1},\{J^r,G^{ra_2}\}\}, & \qquad &
M_{40}^{a_1a_2a_3} = d^{a_2a_3e_1} \{T^{e_1},\{J^r,G^{ra_1}\}\}, \\
M_{41}^{a_1a_2a_3} = f^{a_1e_1g_1} f^{a_2e_1h_1} \{T^{a_3},\{G^{rg_1},G^{rh_1}\}\}, & \qquad &
M_{42}^{a_1a_2a_3} = f^{a_1e_1g_1} f^{a_2e_1h_1} \{T^{g_1},\{G^{ra_3},G^{rh_1}\}\}, \\
M_{43}^{a_1a_2a_3} = f^{a_1e_1g_1} f^{a_2e_1h_1} \{T^{h_1},\{G^{ra_3},G^{rg_1}\}\}, & \qquad &
M_{44}^{a_1a_2a_3} = f^{a_1e_1h_1} f^{a_3e_1g_1} \{T^{a_2},\{G^{rg_1},G^{rh_1}\}\}, \\
M_{45}^{a_1a_2a_3} = f^{a_1e_1h_1} f^{a_3e_1g_1} \{T^{g_1},\{G^{ra_2},G^{rh_1}\}\}, & \qquad &
M_{46}^{a_1a_2a_3} = f^{a_1e_1h_1} f^{a_3e_1g_1} \{T^{h_1},\{G^{ra_2},G^{rg_1}\}\}, \\
M_{47}^{a_1a_2a_3} = f^{a_2e_1g_1} f^{a_3e_1h_1} \{T^{a_1},\{G^{rg_1},G^{rh_1}\}\}, & \qquad &
M_{48}^{a_1a_2a_3} = f^{a_2e_1g_1} f^{a_3e_1h_1} \{T^{g_1},\{G^{ra_1},G^{rh_1}\}\}, \\
M_{49}^{a_1a_2a_3} = f^{a_2e_1g_1} f^{a_3e_1h_1} \{T^{h_1},\{G^{ra_1},G^{rg_1}\}\}, & \qquad &
M_{50}^{a_1a_2a_3} = d^{a_1e_1g_1} d^{a_2e_1h_1} \{T^{h_1},\{G^{ra_3},G^{rg_1}\}\}, \\
M_{51}^{a_1a_2a_3} = i d^{a_1e_1g_1} f^{a_2e_1h_1} \{T^{g_1},\{G^{ra_3},G^{rh_1}\}\}, & \qquad &
M_{52}^{a_1a_2a_3} = i d^{a_1e_1g_1} f^{a_2e_1h_1} \{T^{h_1},\{G^{ra_3},G^{rg_1}\}\}, \\
M_{53}^{a_1a_2a_3} = i f^{a_1e_1h_1} d^{a_3e_1g_1} \{T^{g_1},\{G^{ra_2},G^{rh_1}\}\}, & \qquad &
M_{54}^{a_1a_2a_3} = i f^{a_1e_1h_1} d^{a_3e_1g_1} \{T^{h_1},\{G^{ra_2},G^{rg_1}\}\}, \\
M_{55}^{a_1a_2a_3} = i d^{a_2e_1g_1} f^{a_3e_1h_1} \{T^{g_1},\{G^{ra_1},G^{rh_1}\}\}, & \qquad &
M_{56}^{a_1a_2a_3} = i d^{a_2e_1g_1} f^{a_3e_1h_1} \{T^{h_1},\{G^{ra_1},G^{rg_1}\}\}, \\
M_{57}^{a_1a_2a_3} = i f^{a_1e_1h_1} d^{a_2e_1g_1} \{T^{h_1},\{G^{ra_3},G^{rg_1}\}\}, & \qquad &
M_{58}^{a_1a_2a_3} = i f^{a_2e_1h_1} d^{a_3e_1g_1} \{T^{h_1},\{G^{ra_1},G^{rg_1}\}\}, \\
M_{59}^{a_1a_2a_3} = i d^{a_1e_1g_1} f^{a_3e_1h_1} \{T^{h_1},\{G^{ra_2},G^{rg_1}\}\}. & \qquad &
\end{array}
\end{eqnarray}

\subsubsection{${\sf M}^{a_1a_2}$ basis}

The ${\sf M}^{a_1a_2}$ basis can now be obtained from ${\sf M}^{a_1a_2a_3}$ by contracting two flavor indices with $if^{a_1a_2a_3}$ or $d^{a_1a_2a_3}$. In either case, the procedure yields,
\begin{equation}
{\sf M}^{a_1a_2} = \{M_i^{a_1a_2}\}, \label{eq:mab}
\end{equation}
where
\begin{eqnarray}
\begin{array}{lll}
M_{1}^{a_1a_2} = \delta^{a_1a_2}, & \qquad &
M_{2}^{a_1a_2} = i f^{a_1a_2e_1} T^{e_1}, \\
M_{3}^{a_1a_2} = d^{a_1a_2e_1} T^{e_1}, & \qquad &
M_{4}^{a_1a_2} = \delta^{a_1a_2} J^2, \\
M_{5}^{a_1a_2} = \{T^{a_1},T^{a_2}\}, & \qquad &
M_{6}^{a_1a_2} = \{G^{ra_1},G^{ra_2}\}, \\
M_{7}^{a_1a_2} = d^{a_1a_2e_1} \{J^r,G^{re_1}\}, & \qquad &
M_{8}^{a_1a_2} = i f^{a_1a_2e_1} \{J^r,G^{re_1}\}, \\
M_{9}^{a_1a_2} = \{T^{a_1},\{J^r,G^{ra_2}\}\}, & \qquad &
M_{10}^{a_1a_2} = \{T^{a_2},\{J^r,G^{ra_1}\}\}, \\
M_{11}^{a_1a_2} = i f^{a_1a_2e_1} \{J^2,T^{e_1}\}, & \qquad &
M_{12}^{a_1a_2} = d^{a_1a_2e_1} \{J^2,T^{e_1}\}.
\end{array}
\end{eqnarray}

\subsubsection{${\sf M}^{a_1}$ basis}

The ${\sf M}^{a_1}$ basis can be obtained simply as $\delta^{a_2a_3}{\sf M}^{a_1a_2a_3}$, $if^{a_1a_2a_3}{\sf M}^{a_2a_3}$ or $d^{a_1a_2a_3}{\sf M}^{a_2a_3}$. The resultant basis reads,
\begin{equation}
{\sf M}^{a} = \{M_i^{a_1}\}, \label{eq:ma}
\end{equation}
where
\begin{equation}
M_{1}^{a_1} = T^{a_1}, \qquad
M_{2}^{a_1} = \{J^r,G^{ra_1}\}, \qquad
M_{3}^{a_1} = \{J^2,T^{a_1}\}.
\end{equation}

\subsubsection{${\sf M}$ basis}

There are several ways to obtain the ${\sf M}$ basis: $if^{a_1a_2a_3} {\sf M}^{a_1a_2a_3}$, $d^{a_1a_2a_3}{\sf M}^{a_1a_2a_3}$, $\delta^{a_1a_2}{\sf M}^{a_1a_2}$, or
any product that saturates the flavor indices, for instance, the tensor product of $T^{a_1}$ with the operators in the ${\sf M}^{a_1}$ basis, as long as up to 3-body operators are retained. The resultant basis, after removing redundant operators and/or irrelevant constant factors, is
\begin{equation}
{\sf M} = \{M_i\}, \label{eq:mzero}
\end{equation}
with
\begin{equation}
M_1 = \openone, \qquad M_2 = J^2, \label{eq:m}
\end{equation}
which of course reduces to the operator basis used to construct the $1/N_c$ expansion of the baryon mass operator in the flavor symmetry limit (\ref{eq:su3mass}).

\subsection{Flavor projection operators}

Once the bases ${\sf M}$, ${\sf M}^{a_1}$, ${\sf M}^{a_1a_2}$ and ${\sf M}^{a_1a_2a_3}$ are defined, the next step is to set a mechanism to manipulate them according to their transformation properties under decompositions (\ref{eq:8x8}) and (\ref{eq:8x8x8}). A suitable method is the one based on the operator projection technique of Refs.~\cite{banda1,banda2}. This technique uses the decomposition of the tensor space formed by the product of the adjoint space with itself $n$ times, $\prod_{i = 1}^n adj \otimes$, into subspaces labeled by a specific eigenvalue of the quadratic Casimir operator $C$ of $SU(3)$. The projection operators $\pr{m}$ that can be constructed for each subspace read,
\begin{equation}
\pr{m} = \prod_{i = 1}^k \left[ \frac{C - c_{n_i}\openone}{c_{m} - c_{n_i}} \right], \qquad \qquad c_m \neq c_{n_i}, \label{eq:general_proj}
\end{equation}
where $k$ labels the number of different possible eigenvalues for $C$ and $c_{m}$ are its eigenvalues given by
\begin{equation}
c_m = \frac12 \left[n N - \frac{n^2}{N} + \sum_i r_i^2 - \sum_i c^2_i \right], \label{eq:egenvalues_C}
\end{equation}
where $n$ is the total number of boxes of the Young tableu for a given representation, $r_i$ is the number of boxes in the $i$th row and $c_i$ is the number of boxes in the $i$th column \cite{Gross}.

Thus, for the product of two $SU(3)$ adjoints, the flavor projectors $[\pr{m}]^{a_1a_2a_3a_4}$ for the irreducible representation of dimension $m$ contained in (\ref{eq:8x8}) read \cite{banda1},
\begin{equation}
[\pr{1}]^{a_1a_2a_3a_4} = \frac{1}{N_f^2-1} \delta^{a_1a_2} \delta^{a_3a_4}, \label{eq:p1ab}
\end{equation}
\begin{equation}
[\pr{8}]^{a_1a_2a_3a_4} = \frac{N_f}{N_f^2-4} d^{a_1a_2e_1} d^{a_3a_4e_1}, \label{eq:p8ab}
\end{equation}
\begin{equation}
[\pr{8_A}]^{a_1a_2a_3a_4} = \frac{1}{N_f} f^{a_1a_2e_1} f^{a_3a_4e_1}, \label{eq:p8aab}
\end{equation}
\begin{equation}
[\pr{10+\overline{10}}]^{a_1a_2a_3a_4} = \frac12 (\delta^{a_1a_3} \delta^{a_2a_4} - \delta^{a_2a_3} \delta^{a_1a_4}) - \frac{1}{N_f} f^{a_1a_2e_1} f^{a_3a_4e_1}, \label{eq:p10ab}
\end{equation}
and
\begin{equation}
[\pr{27}]^{a_1a_2a_3a_4} = \frac12 (\delta^{a_1a_3} \delta^{a_2a_4} + \delta^{a_2a_3} \delta^{a_1a_4}) - \frac{1}{N_f^2-1} \delta^{a_1a_2} \delta^{a_3a_4} - \frac{N_f}{N_f^2-4} d^{a_1a_2e_1} d^{a_3a_4e_1}, \label{eq:p27ab}
\end{equation}
which satisfy the completeness relation
\begin{equation}
[\mathcal{P}^{(1)} + \mathcal{P}^{(8)} + \mathcal{P}^{(8_A)} + \mathcal{P}^{(10+\overline{10})} + \mathcal{P}^{(27)}]^{a_1a_2a_3a_4} = \delta^{a_1a_3} \delta^{a_2a_4}.
\end{equation}

As for the product of three adjoints, following decomposition (\ref{eq:8x8x8}), the projection operators can be constructed as \cite{banda2}
\begin{equation}
[\tilde{\mathcal{P}}^{(m)}]^{a_1a_2a_3b_1b_2b_3} = \left[ \left( \frac{C-c_{n_1} \openone}{c_m-c_{n_1}} \right) \left( \frac{C-c_{n_2} \openone}{c_m-c_{n_2}} \right) \left( \frac{C-c_{n_3} \openone}{c_m-c_{n_3}} \right) \left(\frac{C-c_{n_4} \openone}{c_m-c_{n_4}} \right)\left( \frac{C-c_{n_5} \openone}{c_m-c_{n_5}} \right) \right]^{a_1a_2a_3b_1b_2b_3}, \label{eq:pa_1a_2a_3}
\end{equation}
where $m$ labels the flavor representation of each projector and $n_i$ label flavor representations other than $m$. The Casimir operator can be expressed as
\begin{equation}
[C]^{a_1a_2a_3b_1b_2b_3} = 6 \delta^{a_1b_1} \delta^{a_2b_2} \delta^{a_3b_3} - 2 \delta^{a_1b_1} f^{a_2b_2e_1} f^{a_3b_3e_1} - 2 \delta^{a_2b_2} f^{a_1b_1e_1} f^{a_3b_3e_1} - 2 \delta^{a_3b_3} f^{a_1b_1e_1} f^{a_2b_2e_1},	
\end{equation}
where
\begin{equation}
c_1 = 0, \hspace{0.5cm} c_8 = 3, \hspace{0.5cm} c_{10+\overline{10}} = 6, \hspace{0.5cm} c_{27} = 8, \hspace{0.5cm} c_{35+\overline{35}} = 12, \hspace{0.5cm} c_{64} = 15,
\end{equation}
are its corresponding eigenvalues.

As it was discussed in Ref.~\cite{banda2}, the explicit analytic construction of $[\tilde{\mathcal{P}}^{(m)}]^{a_1a_2a_3b_1b_2b_3}$ is quite involved, so its matrix version is implemented and used instead. Thus, $[\tilde{\mathcal{P}}^{(m)}]^{a_1a_2a_3b_1b_2b_3}$ is replaced with a well defined $512\times 512$ matrix, ${\sf P}^{(m)}$, where
\begin{equation}
{\sf P}^{(m)} {\sf P}^{(m)} = {\sf P}^{(m)}, \qquad \qquad {\sf P}^{(m)} {\sf P}^{(n)} = 0, \quad n\neq m,
\end{equation}
along with
\begin{equation}
{\sf P}^{(1)} + {\sf P}^{(8)} + {\sf P}^{(10+\overline{10})} + {\sf P}^{(27)} + {\sf P}^{(35+\overline{35})} + {\sf P}^{(64)} = {\sf I}_{512},
\end{equation}
where ${\sf I}_{512}$ represents the identity matrix of order $512$. Further details can be found in Ref.~\cite{banda2}.

\subsection{$1/N_c$ expansion for the baryon mass operator in the $SU(3)$ flavor symmetry limit}

The $1/N_c$ expansion for the baryon mass operator in the $SU(3)$ flavor symmetry limit, denoted here by $M_{SU(3)}$, follows Eq.~(\ref{eq:su3mass}) and is given by
\begin{equation}
M_{SU(3)} = m_1^{1,0} N_c \openone + \frac{1}{N_c} m_2^{1,0} J^2, \label{eq:msu3pr}
\end{equation}
where $m_k^{m,I}$, adopting the notation of Ref.~\cite{jl}, denote undetermined coefficients that accompany the baryon operator $M_k$ from operator basis ${\sf M}$, (\ref{eq:mzero}), which transforms under the $SU(3)$ flavor representation of dimension $m$ and with isospin $I$. Notice that the series has been truncated at $N_c=3$.

\subsection{$1/N_c$ expansion for the baryon mass operator including first-order SB}

The $1/N_c$ expansion for the baryon mass operator including first-order SB, denoted here by $M_{\mathrm{sb}1}^{m,I}$, can easily be constructed using the operator basis ${\sf M}^{a_1}$, Eq.~(\ref{eq:ma}). For $I=0$ and $1$, the expansions can be written respectively as
\begin{equation}
M_{\mathrm{sb}1}^{8,0} = m_1^{8,0} T^8 + \frac{1}{N_c} m_2^{8,0} \{J^r,G^{r8}\} + \frac{1}{N_c^2} m_3^{8,0} \{J^2,T^8\}, \label{eq:nsb0}
\end{equation}
and
\begin{equation}
M_{\mathrm{sb}1}^{8,1} = m_1^{8,1} T^3 + \frac{1}{N_c} m_2^{8,1} \{J^r,G^{r3}\} + \frac{1}{N_c^2} m_3^{8,1} \{J^2,T^3\}. \label{eq:nsb1}
\end{equation}

The matrix elements of the octet operators involved in Eqs.~(\ref{eq:nsb0}) and (\ref{eq:nsb1}) are listed in Table \ref{t:sb1} for the sake of completeness.

\begin{table*}[h]
\caption{\label{t:sb1}Matrix elements of baryon operators contributing to first-order SB to the baryon mass.}
\begin{ruledtabular}
\begin{tabular}{lcccccc}
$B$ & $\langle T^3 \rangle$ & $\langle \{J^r,G^{r3}\} \rangle$ & $ \langle \{J^2,T^3\} \rangle$ & $\displaystyle \frac{1}{\sqrt{3}} \langle T^8 \rangle$ & $\displaystyle \frac{1}{\sqrt{3}} \langle \{J^r,G^{r8}\} \rangle$ & $\displaystyle \frac{1}{\sqrt{3}} \langle \{J^2,T^8\} \rangle$ \\
\hline
$n$ & $-\frac12$ & $-\frac54$ & $-\frac34$ & $\frac12$ & $\frac14$ & $\frac34$ \\
$p$ & $\frac12$ & $\frac54$ & $\frac34$ & $\frac12$ & $\frac14$ & $\frac34$ \\
$\Sigma^+$ & $1$ & $1$ & $\frac32$ & $0$ & $\frac12$ & $0$ \\
$\Sigma^0$ & $0$ & $0$ & $0$ & $0$ & $\frac12$ & $0$ \\
$\Sigma^-$ & $-1$ & $-1$ & $-\frac32$ & $0$ & $\frac12$ & $0$ \\
$\Xi^-$ & $-\frac12$ & $\frac14$ & $-\frac34$ & $-\frac12$ & $-\frac34$ & $-\frac34$ \\
$\Xi^0$ & $\frac12$ & $-\frac14$ & $\frac34$ & $-\frac12$ & $-\frac34$ & $-\frac34$ \\
$\Lambda$ & $0$ & $0$ & $0$ & $0$ & $-\frac12$ & $0$ \\
$\Sigma^0 \Lambda$ & $0$ & $\frac{\sqrt{3}}{2}$ & $0$ & $0$ & $0$ & $0$ \\
$\Delta^{++}$ & $\frac32$ & $\frac{15}{4}$ & $\frac{45}{4}$ & $\frac12$ & $\frac54$ & $\frac{15}{4}$ \\
$\Delta^+$ & $\frac12$ & $\frac54$ & $\frac{15}{4}$ & $\frac12$ & $\frac54$ & $\frac{15}{4}$ \\
$\Delta^0$ & $-\frac12$ & $-\frac54$ & $-\frac{15}{4}$ & $\frac12$ & $\frac54$ & $\frac{15}{4}$ \\
$\Delta^-$ & $-\frac32$ & $-\frac{15}{4}$ & $-\frac{45}{4}$ & $\frac12$ & $\frac54$ & $\frac{15}{4}$ \\
${\Sigma^*}^+$ & $1$ & $\frac52$ & $\frac{15}{2}$ & $0$ & $0$ & $0$ \\
${\Sigma^*}^0$ & $0$ & $0$ & $0$ & $0$ & $0$ & $0$ \\
${\Sigma^*}^-$ & $-1$ & $-\frac52$ & $-\frac{15}{2}$ & $0$ & $0$ & $0$ \\
${\Xi^*}^-$ & $-\frac12$ & $-\frac54$ & $-\frac{15}{4}$ & $-\frac12$ & $-\frac54$ & $-\frac{15}{4}$ \\
${\Xi^*}^0$ & $\frac12$ & $\frac54$ & $\frac{15}{4}$ & $-\frac12$ & $-\frac54$ & $-\frac{15}{4}$ \\
$\Omega^-$ & $0$ & $0$ & $0$ & $-1$ & $-\frac52$ & $-\frac{15}{2}$ \\
\end{tabular}
\end{ruledtabular}
\end{table*}

\subsection{$1/N_c$ expansion for the baryon mass operator including second-order SB}

The $1/N_c$ expansion for the baryon mass including second-order SB can directly be obtained from the operator basis ${\sf M}^{a_1a_2}$, Eq.~(\ref{eq:mab}). The most general $1/N_c$ expansion, retaining up to $3$-body operators, reads
\begin{equation}
M_{\mathrm{sb}2}^{a_1a_2} = n_1 N_c M_1^{a_1a_2} + \sum_{k=2}^3 n_k M_k^{a_1a_2} + \frac{1}{N_c} \sum_{k=4}^8 n_k M_k^{a_1a_2} + \frac{1}{N_c^2} \sum_{k=9}^{12} n_k M_k^{a_1a_2}, \label{eq:m2ab}
\end{equation}
where $n_k$ $(k=1,\ldots,12)$ are unknown coefficients. Notice that $M_{\mathrm{sb}2}^{a_1a_2}$ contains components of all allowed flavor representations according to decomposition (\ref{eq:8x8}). A formal way to disentangle them is by using the projection operators (\ref{eq:p1ab})--(\ref{eq:p27ab}). Thus, the $I=0$, $1$, and $2$ pieces of $M_{\mathrm{sb}2}^{a_1a_2}$ are obtained by fixing the two free flavor indices to $\{a_1,a_2\}=\{8,8\}$, $\{a_1,a_2\}=\{3,8\}$, and $\{a_1,a_2\}=\{3,3\}$, respectively. For example, the $I=0$ piece can be written as
\begin{equation}
M_{\mathrm{sb}2}^{m,0} = n_1^{m,0} N_c [\mathcal{P}^{(m)} M_1]^{88} + \sum_{k=2}^3 n_k^{m,0} [\mathcal{P}^{(m)} M_k]^{88} + \frac{1}{N_c} \sum_{k=4}^8 n_k^{m,0} [\mathcal{P}^{(m)} M_k]^{88} + \frac{1}{N_c^2} \sum_{k=9}^{12} n_k^{m,0} [\mathcal{P}^{(m)} M_k]^{88}, \label{eq:m2abI}
\end{equation}
where the dimensions of the allowed $SU(3)$ flavor representations are $m=1$, $8$, $10+\overline{10}$ and $27$. Similar expressions can be obtained for the $I=1$ and $I=2$ pieces.

At this point, the number of free parameters has grown to the extent that the approach seems to have no predictive power. However, here it is where the applicability of projection operators manifests itself to further simplify the analysis.

The action of projections operators on the operator basis ${\sf M}^{a_1a_2}$, Eq.~(\ref{eq:mab}), yields the non-zero structures
\begin{equation}
[\mathcal{P}^{(1)} M_{1}]^{a_1a_2} = \delta^{a_1a_2}, \label{eq:mp1ab}
\end{equation}
\begin{equation}
[\mathcal{P}^{(1)} M_{4}]^{a_1a_2} = \delta^{a_1a_2} J^2,
\end{equation}
\begin{equation}
[\mathcal{P}^{(1)} M_{5}]^{a_1a_2} = \frac{N_c(N_c+2N_f)(N_f-2)}{2N_f(N_f^2-1)} \delta^{a_1a_2} + \frac{2}{N_f^2-1} \delta^{a_1a_2} J^2,
\end{equation}
\begin{equation}
[\mathcal{P}^{(1)} M_{6}]^{a_1a_2} = \frac38 \frac{N_c(N_c+2N_f)}{N_f^2-1} \delta^{a_1a_2} -\frac{N_f+2}{2N_f(N_f^2-1)} \delta^{a_1a_2} J^2,
\end{equation}
\begin{equation}
[\mathcal{P}^{(1)} M_{9}]^{a_1a_2} = \frac{2(N_c+N_f)}{N_f(N_f+1)} \delta^{a_1a_2} J^2,
\end{equation}
\begin{equation}
[\mathcal{P}^{(1)} M_{10}]^{a_1a_2} = \frac{2(N_c+N_f)}{N_f(N_f+1)} \delta^{a_1a_2} J^2,
\end{equation}
\begin{equation}
[\mathcal{P}^{(8)} M_{3}]^{a_1a_2} = d^{a_1a_2e_1} T^{e_1},
\end{equation}
\begin{equation}
[\mathcal{P}^{(8)} M_{5}]^{a_1a_2} = \frac{(N_c+N_f)(N_f-4)}{N_f^2-4} d^{a_1a_2e_1} T^{e_1} + \frac{2N_f}{N_f^2-4} d^{a_1a_2e_1} \{J^r,G^{re_1}\},
\end{equation}
\begin{equation}
[\mathcal{P}^{(8)} M_{6}]^{a_1a_2} = \frac34 \frac{(N_c+N_f)N_f}{N_f^2-4} d^{a_1a_2e_1} T^{e_1} - \frac12 \frac{N_f+4}{N_f^2-4} d^{a_1a_2e_1} \{J^r,G^{re_1}\},
\end{equation}
\begin{equation}
[\mathcal{P}^{(8)} M_{7}]^{a_1a_2} = d^{a_1a_2e_1} \{J^r,G^{re_1}\},
\end{equation}
\begin{equation}
[\mathcal{P}^{(8)} M_{9}]^{a_1a_2} = \frac{N_c+N_f }{N_f+2} d^{a_1a_2e_1} \{J^r,G^{re_1}\} + \frac{1}{N_f+2} d^{a_1a_2e_1} \{J^2,T^{e_1}\},
\end{equation}
\begin{equation}
[\mathcal{P}^{(8)} M_{10}]^{a_1a_2} = \frac{N_c+N_f}{N_f+2} d^{a_1a_2e_1} \{J^r,G^{re_1}\} + \frac{1}{N_f+2} d^{a_1a_2e_1} \{J^2,T^{e_1}\},
\end{equation}
\begin{equation}
[\mathcal{P}^{(8)} M_{12}]^{a_1a_2} = d^{a_1a_2e_1} \{J^2,T^{e_1}\},
\end{equation}
\begin{equation}
[\mathcal{P}^{(8_A)} M_{2}]^{a_1a_2} = i f^{a_1a_2e_1} T^{e_1},
\end{equation}
\begin{equation}
[\mathcal{P}^{(8_A)} M_{8}]^{a_1a_2} = i f^{a_1a_2e_1} \{J^r,G^{re_1}\},
\end{equation}
\begin{equation}
[\mathcal{P}^{(8_A)} M_{11}]^{a_1a_2} = i f^{a_1a_2e_1} \{J^2,T^{e_1}\},
\end{equation}
\begin{equation}
[\mathcal{P}^{(10+\overline{10})} M_{9}]^{a_1a_2} = \frac12 \{T^{a_1},\{J^r,G^{ra_2}\}\} - \frac12 \{T^{a_2},\{J^r,G^{ra_1}\}\},
\end{equation}
\begin{equation}
[\mathcal{P}^{(10+\overline{10})} M_{10}]^{a_1a_2} = -\frac12 \{T^{a_1},\{J^r,G^{ra_2}\}\} + \frac12 \{T^{a_2},\{J^r,G^{ra_1}\}\},
\end{equation}
\begin{eqnarray}
[\mathcal{P}^{(27)} M_{5}]^{a_1a_2} & = & \{T^{a_1},T^{a_2}\} - \frac{N_c(N_c+2N_f)(N_f-2)}{2N_f(N_f^2-1)} \delta^{a_1a_2} - \frac{2}{N_f^2-1} \delta^{a_1a_2} J^2 \nonumber \\
& & \mbox{} - \frac{(N_c+N_f)(N_f-4)}{N_f^2-4} d^{a_1a_2e_1} T^{e_1} - \frac{2N_f}{N_f^2-4} d^{a_1a_2e_1} \{J^r,G^{re_1}\},
\end{eqnarray}
\begin{eqnarray}
[\mathcal{P}^{(27)} M_{6}]^{a_1a_2} & = & \{G^{ra_1},G^{ra_2}\} - \frac38 \frac{N_c(N_c+2N_f)}{N_f^2-1} \delta^{a_1a_2} + \frac12 \frac{N_f+2}{N_f(N_f^2-1)} \delta^{a_1a_2} J^2 \nonumber \\
& & \mbox{} - \frac34 \frac{(N_c+N_f)N_f}{N_f^2-4} d^{a_1a_2e_1} T^{e_1} + \frac12 \frac{N_f+4}{N_f^2-4} d^{a_1a_2e_1} \{J^r,G^{re_1}\},
\end{eqnarray}
\begin{eqnarray}
[\mathcal{P}^{(27)} M_{9}]^{a_1a_2} & = & \frac12 \{T^{a_1},\{J^r,G^{ra_2}\}\} + \frac12 \{T^{a_2},\{J^r,G^{ra_1}\}\} - \frac{2(N_c+N_f)}{N_f(N_f+1)} \delta^{a_1a_2} J^2 \nonumber \\
& & \mbox{} - \frac{N_c+N_f}{N_f+2}d^{a_1a_2e_1} \{J^r,G^{re_1}\} - \frac{1}{N_f+2}d^{a_1a_2e_1} \{J^2,T^{e_1}\},
\end{eqnarray}
\begin{eqnarray}
[\mathcal{P}^{(27)} M_{10}]^{a_1a_2} & = & \frac12 \{T^{a_1},\{J^r,G^{ra_2}\}\} + \frac12 \{T^{a_2},\{J^r,G^{ra_1}\}\} - \frac{2(N_c+N_f)}{N_f(N_f+1)} \delta^{a_1a_2} J^2 \nonumber \\
& & \mbox{} - \frac{N_c+N_f}{N_f+2} d^{a_1a_2e_1} \{J^r,G^{re_1}\} - \frac{1}{N_f+2}d^{a_1a_2e_1} \{J^2,T^{e_1}\}. \label{eq:mp27ab}
\end{eqnarray}

A close inspection to Eqs.~(\ref{eq:p1ab})-(\ref{eq:p27ab}) reveals that all operator coefficients in the singlet and octet representations\footnote{Actually, the antisymmetric octet representation does not contribute to any baryon masses.}, can be reabsorbed into the already existing operator coefficients of Eqs.~(\ref{eq:msu3pr}) and (\ref{eq:nsb0}) and (\ref{eq:nsb1}), respectively. As for the $10+\overline{10}$ representation, only the $I=1$ piece $\{T^{3},\{J^r,G^{r8}\}\} - \{T^{8},\{J^r,G^{r3}\}\}$ contributes with a single coefficient. Finally, for the $27$ representation, only two operators are relevant, namely, the $2$-body operator $\{T^{a_1},T^{a_2}\}$ and the $3$-body operator $\{T^{a_1},\{J^r,G^{ra_2}\}\} + \{T^{a_2},\{J^r,G^{ra_1}\}\}$. The additional $2$-body operator $\{G^{ra_1},G^{ra_2}\}$ can be related to $\{T^{a_1},T^{a_2}\}$ through the identity $4[\mathcal{P}^{(27)}]^{a_1a_2b_1b_2}\{G^{rb_1},G^{rb_2}\}=[\mathcal{P}^{(27)}]^{a_1a_2b_1b_2}\{T^{b_1},T^{b_2}\}$, according to the identities listed in Table VIII of Ref.~\cite{djm95}.

The analysis presented so far agrees in full with the one, at the same order, contained in Ref.~\cite{jl}. Following the notation of this reference, the baryon mass operator can be expressed as
\begin{equation}
M = \sum_{m,I} M^{m,I}, \label{eq:massr}
\end{equation}
where $m$ denotes the relevant $SU(3)$ dimension and $I$ denotes the isospin. Thus, the expressions read \cite{jl}
\begin{equation}
M^{1,0} = \tilde{m}_1^{1,0} N_c \openone + \frac{1}{N_c} \tilde{m}_2^{1,0} J^2, \label{eq:e01}
\end{equation}
\begin{equation}
M^{8,0} = \tilde{m}_1^{8,0} T^8 + \frac{1}{N_c} \tilde{m}_2^{8,0} \{J^r,G^{r8}\} + \frac{1}{N_c^2} \tilde{m}_3^{8,0} \{J^2,T^8\}, \label{eq:e80}
\end{equation}
\begin{equation}
M^{27,0} = \frac{1}{N_c} \tilde{m}_1^{27,0} \{T^8,T^8\} + \frac{1}{N_c^2} \tilde{m}_2^{27,0} \{T^8,\{J^r,G^{r8}\}\}, \label{eq:e270}
\end{equation}
\begin{equation}
M^{8,1} = \tilde{m}_1^{8,1} T^3 + \frac{1}{N_c} \tilde{m}_2^{8,1} \{J^r,G^{r3}\} + \frac{1}{N_c^2} \tilde{m}_3^{8,1} \{J^2,T^3\}. \label{eq:e81}
\end{equation}
\begin{equation}
M^{27,1} = \frac{1}{N_c} \tilde{m}_1^{27,1} \{T^3,T^8\} + \frac{1}{N_c^2} \tilde{m}_2^{27,1} (\{T^3,\{J^r,G^{r8}\}\}\} + \{T^8,\{J^r,G^{r3}\}\}), \label{eq:e271}
\end{equation}
\begin{equation}
M^{10+\overline{10},1} = \frac{1}{N_c^2} \tilde{m}_1^{10+\overline{10},1} (\{T^3,\{J^r,G^{r8}\}\} - \{T^8,\{J^r,G^{r3}\}\}, \label{eq:e101}
\end{equation}
and
\begin{equation}
M^{27,2} = \frac{1}{N_c} \tilde{m}_5^{27,2} \{T^3,T^3\} + \frac{1}{N_c^2} \tilde{m}_1^{27,2} \{T^3,\{J^r,G^{r3}\}\}. \label{eq:e272}
\end{equation}
The first term in Eq.~(\ref{eq:e01}) is the overall spin-independent mass and is common to both baryon octet and decuplet. The spin-dependent term, truncated at $N_c=3$, defines $M_\mathrm{hyperfine}$, which describes the spin splittings of the baryon multiplets \cite{jen96}.

One is thus left with only 15 free effective parameters $\tilde{m}_k^{m,I}$ at this order. Because the effective coefficients are the ones that can be determined, their explicit forms in terms of the original ones are unnecessary.

\subsection{$1/N_c$ expansion for the baryon mass operator including third-order SB}

In a complete parallelism to the previous case, the $1/N_c$ expansion for the baryon mass containing third-order SB, using the baryon operators of the ${\sf M}^{a_1a_2a_3}$ basis, Eq.~(\ref{eq:mabc}), can be expressed as
\begin{equation}
M_{\mathrm{sb}3}^{a_1a_2a_3} = N_c \sum_{k=1}^2 o_k M_k^{a_1a_2a_3} + \sum_{k=3}^5 o_k M_k^{a_1a_2a_3} + \frac{1}{N_c} \sum_{k=6}^{20} o_k M_k^{a_1a_2a_3} + \frac{1}{N_c^2} \sum_{k=21}^{59} o_k M_k^{a_1a_2a_3}, \label{eq:m3abc}
\end{equation}
where $o_k$ $(k=1,\ldots,59)$ are unknown coefficients. $M_{\mathrm{sb}3}^{a_1a_2a_3}$ contains components of all allowed flavor representations $m$ according to decomposition (\ref{eq:8x8x8}). The use of the projection operators $[\tilde{\mathcal{P}}^{(m)}]^{a_1a_2a_3b_1b_2b_3}$ will effectively separate these representations. Now, the $I=0$, $1$, $2$ and $3$ pieces of $M_{\mathrm{sb}3}^{a_1a_2a_3}$ are obtained by fixing the three free flavor indices to $\{a_1,a_2,a_3\} = \{8,8,8\}$, $\{a_1,a_2,a_3\} = \{3,8,8\}$, $\{a_1,a_2,a_3\} = \{3,3,8\}$ and $\{a_1,a_2,a_3\} = \{3,3,3\}$, respectively. The expression of the $I=3$ piece, for instance, reads
\begin{eqnarray}
M_{\mathrm{sb}3}^{m,3} & = & N_c \sum_{k=1}^2 o_k^{m,3} [\tilde{\mathcal{P}}^{(m)} M_k]^{333} + \sum_{k=3}^5 o_k^{m,3} [\tilde{\mathcal{P}}^{(m)} M_k]^{333} + \frac{1}{N_c} \sum_{k=6}^{20} o_k^{m,3} [\tilde{\mathcal{P}}^{(m)} M_k]^{333} \nonumber \\
& & \mbox{} + \frac{1}{N_c^2} \sum_{k=21}^{59} o_k^{m,3} [\tilde{\mathcal{P}}^{(m)} M_k]^{333}, \label{eq:m3abcI}
\end{eqnarray}
where $m=1$, $8$, $10+\overline{10}$, $27$, $35+\overline{35}$ and $64$. The explicit forms of the operator structures $[\tilde{\mathcal{P}}^{(m)} Q_1Q_2Q_3]^{a_1a_2a_3}$, where $Q_j^{a_k}$ are flavor adjoints, are listed in Appendix \ref{app:q1q2q3} for $I=0,\ldots,3$. From these structures, expressions as (\ref{eq:m3abcI}) can straightforwardly be obtained.

In order to construct the full expressions for the baryon masses including third-order SB, the matrix elements of the operators in the basis ${\sf M}^{a_1a_2a_3}$ should be obtained. All these matrix elements are provided in the supplementary material to this paper for $I=0,\ldots,3$ and all participating flavor representations. In particular, it can be confirmed that the $35+\overline{35}$ representation does not contribute neither to octet nor decuplet baryon masses and that the $64$ representation only contributes to decuplet baryon masses.

From these matrix elements, after a thorough analysis, it can be shown that the totality of the operator coefficients which come along the baryon operators in the $1$, $8$, and $27$ flavor representations contained in (\ref{eq:m3abc}) can be absorbed into the already effective operators coefficients introduced in Eqs.~(\ref{eq:e01})-(\ref{eq:e272}). The exceptions come from the $10+\overline{10}$ and $64$ representations.

Operators in the $10+\overline{10}$ representation start contributing to the masses of octet baryons with $3$-body operators. The non-zero contributions from $M_{26}^{a_1a_2a_3}$ to $M_{50}^{a_1a_2a_3}$ with $I=3$ and $I=1$ can be absorbed into the $\tilde{m}_1^{10+\overline{10},1}$ coefficient introduced in Eq.~(\ref{eq:e101}). However, operators $M_{51}^{a_1a_2a_3}$ to $M_{59}^{a_1a_2a_3}$ {\it only} contribute to the off-diagonal mass $\Sigma^0\Lambda$ for $I=3$ and $I=1$. These can be cast into
\begin{equation}
M_\mathrm{sb3}^{10+\overline{10},3} = \frac{1}{N_c^2} \sum_{k=51}^{59} o_k^{10+\overline{10},3} [\tilde{\mathcal{P}}^{(10+\overline{10})} M_k]^{333},
\end{equation}
and
\begin{equation}
M_\mathrm{sb3}^{10+\overline{10},1} = \frac{1}{N_c^2} \sum_{k=51}^{59} o_k^{10+\overline{10},1} [\tilde{\mathcal{P}}^{(10+\overline{10})} M_k]^{388}.
\end{equation}
Because $M_{26}^{a_1a_2a_3}$ to $M_{50}^{a_1a_2a_3}$ constitute a subset of linearly independent operators, there is no reason to rule out their contributions. After evaluating the corresponding matrix elements involved, two effective coefficients $\tilde{m}_2^{10+\overline{10},I}$, for $I=3$ and $I=1$, can be defined, namely,
\begin{equation}
\tilde{m}_2^{10+\overline{10},I} = o^{10+\overline{10},I}_{51} - o^{10+\overline{10},I}_{52} + o^{10+\overline{10},I}_{53} - o^{10+\overline{10},I}_{54} + o^{10+\overline{10},I}_{55} - o^{10+\overline{10},I}_{56} - o^{10+\overline{10},I}_{57} - o^{10+\overline{10},I}_{58} - o^{10+\overline{10},I}_{59}.
\end{equation}
Therefore, the use of flavor projection operators has allowed one to find an extra contribution from the $10+\overline{10}$ representation which is not apparent in the analysis of Ref.~\cite{jl}. Although two effective coefficients are needed to account for its effects, the fact that these coefficients appear in only one mass makes it possible to joint them into a single one in an actual analysis.

As for the $64$ representation, operators $M_{26}^{a_1a_2a_3}$ to $M_{50}^{a_1a_2a_3}$ also constitute a subset of linearly independent operators. In principle, all these operators contribute to the baryon masses alike, so they can not be ruled out. This is not a drawback; as these operators affect only decuplet baryons, they effects can be parameterized in terms of a single coefficient for each $I$, namely,
\begin{eqnarray}
\tilde{m}_1^{64,I} & = & o^{64,I}_{26} + o^{64,I}_{27} + o^{64,I}_{28} + \frac14 o^{64,I}_{29} + \frac14 o^{64,I}_{30} + \frac14 o^{64,I}_{31} - \frac14 o^{64,I}_{41} - \frac14 o^{64,I}_{42} - \frac14 o^{64,I}_{43} - \frac14 o^{64,I}_{44} \nonumber \\
 & & \mbox{} - \frac14 o^{64,I}_{45} - \frac14 o^{64,I}_{46} - \frac14 o^{64,I}_{47} - \frac14 o^{64,I}_{48} - \frac14 o^{64,I}_{49} + \frac{1}{12} o^{64,I}_{50}.
\end{eqnarray}

Thus, one is finally left with 21 unknown effective parameters: 2 parameters from $SU(3)$ symmetric expressions, 6 and 7 parameters from first- and second-order SB, respectively, and 6 additional ones from third-order SB. Apart from the coefficients in the $10+\overline{10}$ representation which only affects off-diagonal mass $\Sigma^0\Lambda$, the analysis is consistent with the one presented in Ref.~\cite{jl}.

The full theoretical expressions for baryon masses are listed in Appendix \ref{app:fullmass} for the sake of completeness.

\section{\label{sec:num}Numerical results}

At this stage, it is possible to produce some numbers through a least-squares fit to data. The aim of this exercise is not to be definite about baryon mass determinations, but rather to test the working assumptions. The available experimental data about baryon masses is listed in the Review of Particle Physics \cite{part}; it comprises the masses of $N$, $\Sigma$, $\Xi$, $\Lambda$, $\Sigma^*$, $\Xi^*$ and $\Omega$, together with several baryon mass differences. To determine the 19 free parameters, which result from omitting momentarily $c_2^{10+\overline{10},3}$ and $c_2^{10+\overline{10},1}$ which come from $M_{\Sigma^0\Lambda}$ and from which there is no information whatsoever, data from lattice analyses can be used. The analysis can be carried out on an equal footing by using the data about leading isospin breaking effects in $N$ and $\Delta$ from Ref.~\cite{romiti}.

The data selected to be used in the fit are the measured masses of baryons along with $M_n-M_p$ reported in Ref.~\cite{part}, which makes 15 pieces of data. From lattice results, the value of $M_{\Delta^-}$ is used, along with 6 mass differences between $\Delta$ baryons from Ref.~\cite{romiti}. The fit can be performed under different assumptions; neglecting for instance all terms suppressed by $1/N_c^2$ would be one choice, neglecting the $64$ representation contributions would be another. However, without further ado, the fit can be performed straightforwardly with the data mentioned above for all 19 parameters. An arbitrary error of $0.50 \,\, \mathrm{MeV}$ is added in quadrature to neutron and proton masses to avoid a bias in favor of these best measured values.

The best-fit parameters produced, in units of MeV, are
\begin{eqnarray}
\label{eq:bestfit}
\begin{array}{lll}
\tilde{m}_1^{1,0} = 363.96 \pm 0.02, \quad \quad &
\tilde{m}_2^{1,0} = 237.31 \pm 0.20, & \\
\tilde{m}_1^{8,0} = -454.85 \pm 0.17, \quad \quad &
\tilde{m}_2^{8,0} = 247.89 \pm 0.14, \quad \quad &
\tilde{m}_3^{8,0} = -42.87 \pm 0.74, \\
\tilde{m}_1^{27,0} = -23.48 \pm 0.14, \quad \quad &
\tilde{m}_2^{27,0} = 31.36 \pm.39, \quad \quad & \\
\tilde{m}_1^{64,0} = 22.63 \pm 1.95, \quad \quad & \quad \quad & \\
\tilde{m}_1^{8,1} = -4.79 \pm 0.10, \quad \quad &
\tilde{m}_2^{8,1} = 0.95 \pm 0.36, \quad \quad &
\tilde{m}_3^{8,1} = 2.53 \pm 0.30, \\
\tilde{m}_1^{10+\overline{10},1} = 0.19 \pm 0.60, \quad \quad & \quad \quad & \\
\tilde{m}_1^{27,1} = 27.27 \pm 2.28, \quad \quad & 
\tilde{m}_2^{27,1} = -29.99 \pm 2.94, \quad \quad & \\
\tilde{m}_1^{64,1} = -0.99 \pm 1.56, \quad \quad & \quad \quad & \\
\tilde{m}_1^{27,2} = 1.23 \pm 0.18, \quad \quad &
\tilde{m}_2^{27,2} = -0.23 \pm 0.44, \quad \quad & \\
\tilde{m}_1^{64,2} = -3.27 \pm 3.99 \quad \quad & \quad \quad & \\
\tilde{m}_1^{64,3} = -0.04 \pm 0.20, \quad \quad & \quad \quad & 
\end{array}
\end{eqnarray}
with $\chi^2 = 0.05$ for two degrees of freedom. The rather low value of $\chi^2$ is nothing but a consequence of the working assumptions. The errors indicated in the best-fit parameters come from the fit only and do not include any theoretical uncertainties. According to expectations, the best-fit parameters roughly follow the natural suppression in $1/N_c$, i.e., the leading-order coefficients $\tilde{m}_1^{1,0}$ and $\tilde{m}_2^{1,0}$ coming from the singlet are the most significant ones, followed by the coefficients from the $8$, $10+\overline{10}$, $27$ and $64$ representations, which tend to be less significant. For the latter, the errors obtained are systematically comparable to the central value so they are poorly determined.

With the best-fit parameters, the predicted baryons masses $M_B$ are listed in Table \ref{t:mass}, where
\begin{equation}
M_B = \sum_I M_B^I, \label{eq:fullm}
\end{equation}
and $M_B^I$ are the contributions to $M_B$ for $I=0,\ldots,3$, from the different flavor representations $M_B^{m,I}$,
\begin{equation}
M_B^I= \sum_m M_B^{m,I}.
\end{equation}

\begin{turnpage}

\begin{table*}[h]
\caption{\label{t:mass}Mass of baryon $B$, $M_B=\sum_I M_B^I$, and its contributions $M_B^I= \sum_m M_B^{m,I}$ using the best-fit parameters (\ref{eq:bestfit}). Mass values are given in $\mathrm{MeV}$. The entries at the bottom line indicate the naive
symmetry-breaking parameters associated with $M_B^{m,I}$ at leading order in SB \cite{jl}.}
\begin{ruledtabular}
\begin{tabular}{lrrrrrrrrrrrrrrrr}
$B$ & $M_B$ & $M_B^0$ & $M_B^{1,0}$ & $M_B^{8,0}$ & $M_B^{27,0}$ & $M_B^{64,0}$ & $M_B^1$ & $M_B^{8,1}$ & $M_B^{10+\overline{10},1}$ & $M_B^{27,1}$ & $M_B^{64,1}$ & $M_B^2$ & $M_B^{27,2}$ & $M_B^{64,2}$ & $M_B^{64,3}$ \\
\hline
$n$               & $939.580$ & $938.914$ & $1151.207$ & $-210.339$ & $-1.954$ & $0.000$ & $0.647$ & $1.787$ & $0.011$ & $-1.151$ & $0.000$ & $0.019$ & $0.019$ & $0.000$ & $0.000$ \\
$p$               & $938.287$ & $938.914$ & $1151.207$ & $-210.339$ & $-1.954$ & $0.000$ & $-0.647$ & $-1.787$ & $-0.011$ & $1.151$ & $0.000$ & $0.019$ & $0.019$ & $0.000$ & $0.000$ \\
$\Sigma^+$        & $1189.384$ & $1193.174$ & $1151.207$ & $41.315$ & $0.651$ & $0.000$ & $-4.039$ & $-4.050$ & $0.011$ & $0.000$ & $0.000$ & $0.249$ & $0.249$ & $0.000$ & $0.000$ \\
$\Sigma^0$        & $1192.656$ & $1193.174$ & $1151.207$ & $41.315$ & $0.651$ & $0.000$ & $0.000$ & $0.000$ & $0.000$ & $0.000$ & $0.000$ & $-0.518$ & $-0.518$ & $0.000$ & $0.000$ \\
$\Sigma^-$        & $1197.463$ & $1193.174$ & $1151.207$ & $41.315$ & $0.651$ & $0.000$ & $4.039$ & $4.050$ & $-0.011$ & $0.000$ & $0.000$ & $0.249$ & $0.249$ & $0.000$ & $0.000$ \\
$\Xi^-$           & $1321.722$ & $1318.278$ & $1151.207$ & $169.024$ & $-1.954$ & $0.000$ & $3.425$ & $2.263$ & $0.011$ & $1.151$ & $0.000$ & $0.019$ & $0.019$ & $0.000$ & $0.000$ \\
$\Xi^0$           & $1314.872$ & $1318.278$ & $1151.207$ & $169.024$ & $-1.954$ & $0.000$ & $-3.425$ & $-2.263$ & $-0.011$ & $-1.151$ & $0.000$ & $0.019$ & $0.019$ & $0.000$ & $0.000$ \\
$\Lambda$         & $1115.697$ & $1115.754$ & $1151.207$ & $-41.315$ & $5.862$ & $0.000$ & $0.000$ & $0.000$ & $0.000$ & $0.000$ & $0.000$ & $-0.058$ & $-0.058$ & $0.000$ & $0.000$ \\
$\Sigma^0\Lambda$ & $-1.719$ & $0.000$ & $0.000$ & $0.000$ & $0.000$ & $0.000$ & $-1.719$ & $0.275$ & $0.000$ & $-1.994$ & $0.000$ & $0.000$ & $0.000$ & $0.000$ & $0.000$ \\
$\Delta^{++}$     & $1246.145$ & $1247.963$ & $1388.517$ & $-141.997$ & $0.796$ & $0.647$ & $-2.425$ & $-2.833$ & $0.000$ & $0.455$ & $-0.047$ & $0.612$ & $0.726$ & $-0.114$ & $-0.006$ \\
$\Delta^+$        & $1246.608$ & $1247.963$ & $1388.517$ & $-141.997$ & $0.796$ & $0.647$ & $-0.808$ & $-0.944$ & $0.000$ & $0.152$ & $-0.016$ & $-0.564$ & $-0.657$ & $0.093$ & $0.016$ \\
$\Delta^0$        & $1248.192$ & $1247.963$ & $1388.517$ & $-141.997$ & $0.796$ & $0.647$ & $0.808$ & $0.944$ & $0.000$ & $-0.152$ & $0.016$ & $-0.564$ & $-0.657$ & $0.093$ & $-0.016$ \\
$\Delta^-$        & $1251.006$ & $1247.963$ & $1388.517$ & $-141.997$ & $0.796$ & $0.647$ & $2.425$ & $2.833$ & $0.000$ & $-0.455$ & $0.047$ & $0.612$ & $0.726$ & $-0.114$ & $0.006$ \\
${\Sigma^*}^+$    & $1382.842$ & $1384.604$ & $1388.517$ & $0.000$ & $-1.327$ & $-2.586$ & $-2.186$ & $-1.888$ & $0.000$ & $-0.455$ & $0.158$ & $0.422$ & $0.173$ & $0.249$ & $0.001$ \\
${\Sigma^*}^0$    & $1383.712$ & $1384.604$ & $1388.517$ & $0.000$ & $-1.327$ & $-2.586$ & $0.000$ & $0.000$ & $0.000$ & $0.000$ & $0.000$ & $-0.893$ & $-0.519$ & $-0.374$ & $0.000$ \\
${\Sigma^*}^-$    & $1387.211$ & $1384.604$ & $1388.517$ & $0.000$ & $-1.327$ & $-2.586$ & $2.186$ & $1.888$ & $0.000$ & $0.455$ & $-0.158$ & $0.422$ & $0.173$ & $0.249$ & $-0.001$ \\
${\Xi^*}^-$       & $1535.211$ & $1533.598$ & $1388.517$ & $141.997$ & $-0.796$ & $3.879$ & $1.709$ & $0.944$ & $0.000$ & $0.607$ & $0.158$ & $-0.097$ & $-0.035$ & $-0.062$ & $0.001$ \\
${\Xi^*}^0$       & $1531.790$ & $1533.598$ & $1388.517$ & $141.997$ & $-0.796$ & $3.879$ & $-1.709$ & $-0.944$ & $0.000$ & $-0.607$ & $-0.158$ & $-0.097$ & $-0.035$ & $-0.062$ & $-0.001$ \\
$\Omega^-$        & $1672.458$ & $1672.313$ & $1388.517$ & $283.993$ & $2.388$ & $-2.586$ & $0.000$ & $0.000$ & $0.000$ & $0.000$ & $0.000$ & $0.145$ & $0.104$ & $0.042$ & $0.000$ \\
\hline
 & & & $\mathcal{O}(1)$ & $\mathcal{O}(\epsilon)$ & $\mathcal{O}(\epsilon^2)$ & $\mathcal{O}(\epsilon^3)$ & & $\mathcal{O}(\epsilon^\prime)$ & $\mathcal{O}(\epsilon^\prime\epsilon)$ & $\mathcal{O}(\epsilon^\prime\epsilon)$ & $\mathcal{O}(\epsilon^\prime\epsilon^2)$ & & $\mathcal{O}(\epsilon^{\prime\prime})$ & $\mathcal{O}(\epsilon^{\prime\prime}\epsilon)$ & $\mathcal{O}(\epsilon^{\prime\prime}\epsilon^\prime)$
\end{tabular}
\end{ruledtabular}
\end{table*}

\end{turnpage}

\section{\label{sec:massrel}Baryon mass relations}

A number of interesting relations among baryon masses are obtained by successively neglecting operators in the mass expansion (\ref{eq:fullm}). This was done in detail in Ref.~\cite{jl}, where the isospin sectors $I=0$, $1$, $2$, $3$ were classified. A complete list of those relations is provided in Table II of that reference. In this section, some mass relations are evaluated both analytically and numerically, as an application of the best-fit parameters (\ref{eq:bestfit}).

\subsection{$I = 0$ baryon mass relations}

The mass combinations transforming as $I=0$ are found to be \cite{jl},
\begin{subequations}
\begin{equation}
N_0 = \frac12 (M_n + M_p),
\end{equation}
\begin{equation}
\Sigma_0 = \frac13 (M_{\Sigma^+} + M_{\Sigma^0} + M_{\Sigma^-}),
\end{equation}
\begin{equation}
\Xi_0 = \frac12 (M_{\Xi^-} + M_{\Xi^0}),
\end{equation}
\begin{equation}
\Lambda_0 = M_{\Lambda},
\end{equation}
\begin{equation}
\Delta_0 = \frac14 (M_{\Delta^{++}} + M_{\Delta^+} + M_{\Delta^0} + M_{\Delta^-}),
\end{equation}
\begin{equation}
\Sigma_0^* = \frac13 (M_{{\Sigma^*}^+} + M_{{\Sigma^*}^0} + M_{{\Sigma^*}^-}),
\end{equation}
\begin{equation}
\Xi_0^* = \frac12 (M_{{\Xi^*}^-} + M_{{\Xi^*}^0}).
\end{equation}
\end{subequations}
\begin{equation}
\Omega_0 = M_{\Omega^-}.
\end{equation}

Two well-known mass relations can be tested, namely, the Gell-Mann--Okubo mass relation and the decuplet equal spacing rule. The former can be written as
\begin{eqnarray}
\frac34 \Lambda_0 + \frac14 \Sigma_0 - \frac12 N_0 - \frac12 \Xi_0 & = & - \frac{1}{6N_c} \tilde{m}_1^{27,2} - \frac{1}{6N_c^2} \tilde{m}_2^{27,2} - \frac{3}{2N_c} \tilde{m}_1^{27,0} - \frac{3}{2N_c^2} \tilde{m}_2^{27,0} \nonumber \\
& = & 6.45 \,\, \mathrm{MeV} \label{eq:gmo},
\end{eqnarray}
which is broken by the 27 flavor representation at order $\mathcal{O}(\epsilon^{\prime\prime})$ by the first and second summands and at order $\mathcal{O}(\epsilon^2)$ by the third and fourth summands on the right-hand side.

The equal-spacing rule is usually written as
\begin{equation}
\Delta_0 - \Sigma_0^* = \Sigma_0^* - \Xi_0^* = \Xi_0^* - \Omega_0.
\end{equation}

Two separate relations yield
\begin{equation}
(\Delta_0 - \Sigma_0^*) - (\Sigma_0^* - \Xi_0^*) = \frac{1}{3N_c} \tilde{m}_1^{27,2} + \frac{5}{6N_c^2} \tilde{m}_2^{27,2} + \frac{3}{N_c} \tilde{m}_1^{27,0} + \frac{15}{2N_c^2} \tilde{m}_2^{27,0} + \frac{3}{7N_c^2} \tilde{m}_1^{64,2} + \frac{27}{7N_c^2} \tilde{m}_1^{64,0}, \label{eq:es1} 
\end{equation}
and
\begin{equation}
(\Sigma_0^* - \Xi_0^*) - (\Xi_0^* - \Omega_0) = \frac{1}{3N_c} \tilde{m}_1^{27,2} + \frac{5}{6N_c^2} \tilde{m}_2^{27,2} + \frac{3}{N_c} \tilde{m}_1^{27,0} + \frac{15}{2N_c^2} \tilde{m}_2^{27,0} - \frac{4}{7N_c^2} \tilde{m}_1^{64,2} - \frac{36}{7N_c^2} \tilde{m}_1^{64,0}, \label{eq:es2}
\end{equation}
thus relations (\ref{eq:es1}) and (\ref{eq:es2}) can be combined to get the most highly suppressed operators, which come from the 64 representation. This corresponds to the mass relation [cf.\ Eq.~(4.2) of Ref.~\cite{jl}],
\begin{eqnarray}
\frac12 (\Delta_0 - 3 \Sigma_0^* + 3 \Xi_0^* - \Omega_0) & = & \frac{1}{2N_c^2} \tilde{m}_1^{64,2} + \frac{9}{2N_c^2} \tilde{m}_1^{64,0} \nonumber \\
& = & 11.13 \,\, \mathrm{MeV},
\end{eqnarray}
where the first and second summands on the right-hand side occur at orders $\mathcal{O}(\epsilon^{\prime\prime}\epsilon)$ and $\mathcal{O}(\epsilon^3)$, respectively.

At next subleading order, there is a mass relation given by [cf.\ Eq.~(4.3) of Ref.~\cite{jl}]
\begin{eqnarray}
2\left[ \frac34 \Lambda_0 + \frac14 \Sigma_0 - \frac12 N_0 - \frac12 \Xi_0 \right] + \frac17 (4\Delta_0 - 5\Sigma_0^* - 2\Xi_0^* + 3 \Omega_0) & = & \frac{1}{2N_c^2} \tilde{m}_2^{27,2} + \frac{9}{2N_c^2} \tilde{m}_2^{27,0} \nonumber \\
& = & 15.67 \,\, \mathrm{MeV},
\end{eqnarray}

Another interesting relation which originates from the difference between the average decuplet and octet masses is [cf.\ Eq.~(4.4) of Ref.~\cite{jl}]
\begin{eqnarray}
\frac{1}{10} (4\Delta_0 + 3\Sigma_0^* + 2\Xi_0^* + \Omega_0) - \frac18 (2N_0 + 3\Sigma_0 + \Lambda_0 + 2\Xi_0) & = & \frac{3}{N_c} \tilde{m}_2^{1,0} \nonumber \\
& = & 237.31 \,\, \mathrm{MeV},
\end{eqnarray}
Thus, in the large-$N_c$ limit, the baryon decuplet and baryon octet become degenerate, which is a very well-known result \cite{djm94,djm95}. At $N_c=3$ the above expression defines the average mass difference between decuplet and octet baryons in terms of $\tilde{m}_2^{1,0}$.

\subsection{$I = 1$ baryon mass relations}
The $I=1$ mass combinations can be given as
\begin{equation}
N_1 = - M_n + M_p,
\end{equation}
\begin{equation}
\Sigma_1 = M_{\Sigma^+} - M_{\Sigma^-},
\end{equation}
\begin{equation}
\Xi_1 = - M_{\Xi^-} + M_{\Xi^0},
\end{equation}
\begin{equation}
\Delta_1 = 3 M_{\Delta^{++}} + M_{\Delta^+} - M_{\Delta^0} - 3 M_{\Delta^-},
\end{equation}
\begin{equation}
\Sigma_1^* = M_{{\Sigma^*}^+} - M_{{\Sigma^*}^-},
\end{equation}
\begin{equation}
\Xi_1^* = - M_{{\Xi^*}^-} + M_{{\Xi^*}^0},
\end{equation}
along with the off-diagonal mass $\Sigma^0\Lambda$.

The first test to be performed is $-N_1$, the neutron and proton mass difference; it yields,
\begin{eqnarray}
-N_1 & = & -\tilde{m}_1^{8,1} - \frac{5}{2N_c} \tilde{m}_2^{8,1} - \frac{3}{2N_c^2} \tilde{m}_3^{8,1} + \frac{1}{N_c^2} \tilde{m}_1^{10 + \overline{10},1} - \frac{2}{5N_c} \tilde{m}_1^{27,1} - \frac{2}{5N_c^2} \tilde{m}_2^{27,1} \nonumber \\
& = & (4.79 - 0.79 - 0.42 + 0.02 - 3.64 + 1.33) \,\, \mathrm{MeV} \nonumber \\
& = & 1.29 \,\, \mathrm{MeV}.
\end{eqnarray}
Notice that the smallness of $-N_1$ does not come from the sum of small quantities, but rather from partial cancellations of comparable quantities.

In a similar manner, the most highly suppressed $I=1$ operators in the mass expansion are the ones from the $64$ representation. This leads to the mass relation [cf.\ Eq.~(4.8) of Ref.~\cite{jl}],
\begin{eqnarray}
\Delta_1 - 10\Sigma_1^* + 10\Xi_1^* & = & \frac{12}{N_c^2} \tilde{m}_1^{64,3} + \frac{60}{N_c^2} \tilde{m}_1^{64,1} \nonumber \\
& = & - 6.68 \,\, \mathrm{MeV}, \label{eq:48}
\end{eqnarray}
which gets contributions from orders $\mathcal{O}(\epsilon^{\prime\prime}\epsilon^\prime)$ and $\mathcal{O}(\epsilon^\prime\epsilon^2)$ from the first and second summands on the right-hand side, respectively.

At next order, the Coleman-Glashow relation is obtained,
\begin{eqnarray}
N_1 - \Sigma_1 + \Xi_1 & = & - \frac{3}{N_c^2} \tilde{m}_1^{10 + \overline{10},1}, \nonumber \\
& = & - 0.06 \,\, \mathrm{MeV},
\end{eqnarray}
so violation to this relation comes from the $10+\overline{10}$ representation contribution, which is order $\mathcal{O}(\epsilon\epsilon^\prime)$. Numerically, it is consistent with zero according to its experimental accuracy.

There are three more $I=1$ mass relations listed in Ref.~\cite{jl} [cf.\ Eqs.~(4.10), (4.11), and (4.13) of that reference], namely,
\begin{eqnarray}
N_1 - \Xi_1 + 2\sqrt{3} \Sigma^0\Lambda & = & \frac{6}{N_c} \tilde{m}_2^{8,1} - \frac{2}{5N_c} \tilde{m}_1^{27,1} - \frac{2}{5N_c^2} \tilde{m}_2^{27,1} - \frac{3}{2N_c^2} \tilde{m}_2^{10 + \overline{10},3} + \frac{3}{2N_c^2} \tilde{m}_2^{10 + \overline{10},1}, \nonumber \\
& = & 11.51 \,\, \mathrm{MeV} + \frac16 {\tilde m}_2^{10 + \overline{10},3} - \frac16 {\tilde m}_2^{10 + \overline{10},1}, \label{eq:uk}
\end{eqnarray}
\begin{eqnarray}
\Delta_1 - 3\Sigma_1^* - 4\Xi_1^* & = & \frac{14}{N_c} \tilde{m}_1^{27,1} + \frac{35}{N_c^2} \tilde{m}_2^{27,1} \nonumber \\
& = & 10.62 \,\, \mathrm{MeV}, \label{eq:411}
\end{eqnarray}
and
\begin{eqnarray}
\Sigma_1^* - 2 \Xi_1^* & = & \frac{2}{N_c} \tilde{m}_1^{27,1} + \frac{5}{N_c^2} \tilde{m}_2^{27,1} - \frac{12}{7N_c^2} \tilde{m}_1^{64,3} - \frac{60}{7N_c^2} \tilde{m}_1^{64,1} \nonumber \\
& = & 2.47 \,\, \mathrm{MeV}. \label{eq:413}
\end{eqnarray}

Equation (\ref{eq:uk}) depends on the coefficients $\tilde{m}_2^{10 + \overline{10},3}$ and $\tilde{m}_2^{10 + \overline{10},1}$,  
which remain unknown unless there is a piece of information which allows one to constrain them. Notice that Eqs.~(\ref{eq:411}) and (\ref{eq:413}) can be combined to get Eq.~(\ref{eq:48}) so they are not really independent.

\subsection{$I = 2$ baryon mass relations}

As for $I=2$, there are three mass splittings, namely \cite{jl},
\begin{equation}
\Sigma_2 = M_{\Sigma^+} - 2 M_{\Sigma^0} + M_{\Sigma^-},
\end{equation}
\begin{equation}
\Delta_2 = M_{\Delta^{++}} - M_{\Delta^+} - M_{\Delta^0} + M_{\Delta^-},
\end{equation}
\begin{equation}
\Sigma_2^* = M_{{\Sigma^*}^+} - 2 M_{{\Sigma^*}^0} + M_{{\Sigma^*}^-},
\end{equation}

Direct evaluation of the above expressions leads to
\begin{eqnarray}
\Sigma_2 & = & \frac{4}{N_c} \tilde{m}_1^{27,2} + \frac{4}{N_c^2} \tilde{m}_2^{27,2} \nonumber \\
& = & 1.53 \,\, \mathrm{MeV},
\end{eqnarray}
\begin{eqnarray}
\Delta_2 & = & \frac{8}{N_c} \tilde{m}_1^{27,2} + \frac{20}{N_c^2} \tilde{m}_2^{27,2} + \frac{8}{7N_c^2} \tilde{m}_1^{64,2} \nonumber \\
& = & 2.35 \,\, \mathrm{MeV},
\end{eqnarray}
and
\begin{eqnarray}
\Sigma_2^* & = & \frac{4}{N_c} \tilde{m}_1^{27,2} + \frac{10}{N_c^2} \tilde{m}_2^{27,2} - \frac{24}{7N_c^2} \tilde{m}_1^{64,2} \nonumber \\
& = & 2.63 \,\, \mathrm{MeV}.
\end{eqnarray}

The most highly suppressed relation that can be obtained is [cf.\ Eq.~(4.15) of Ref.~\cite{jl}]
\begin{eqnarray}
\Delta_2 - 2 \Sigma_2^* & = & \frac{8}{N_c^2} \tilde{m}_1^{64,2} \nonumber \\
& = & - 2.91 \,\, \mathrm{MeV},
\end{eqnarray}
which is order $\mathcal{O}(\epsilon^{\prime\prime}\epsilon)$.

\subsection{$I = 3$ baryon mass relations}

There is a single mass relation for $I=3$, which reads [cf.\ Eq.~(4.2) of Ref.~\cite{jl}],
\begin{equation}
\Delta_3 = M_{\Delta^{++}} - 3 M_{\Delta^+} + 3 M_{\Delta^0} - M_{\Delta^-}.
\end{equation}

Straight evaluation of this equation yields
\begin{eqnarray}
\Delta_3 & = & \frac{24}{N_c^2} \tilde{m}_1^{64,3} \nonumber \\
& = & - 0.11 \,\, \mathrm{MeV},
\end{eqnarray}
which occurs at order $\mathcal{O}(\epsilon^{\prime\prime}\epsilon^\prime)$.

The mass relations tested in this section are in good agreement with the $1/N_c$ expectations and, numerically, are well satisfied.

\begin{figure}[ht]
\scalebox{0.30}{\includegraphics{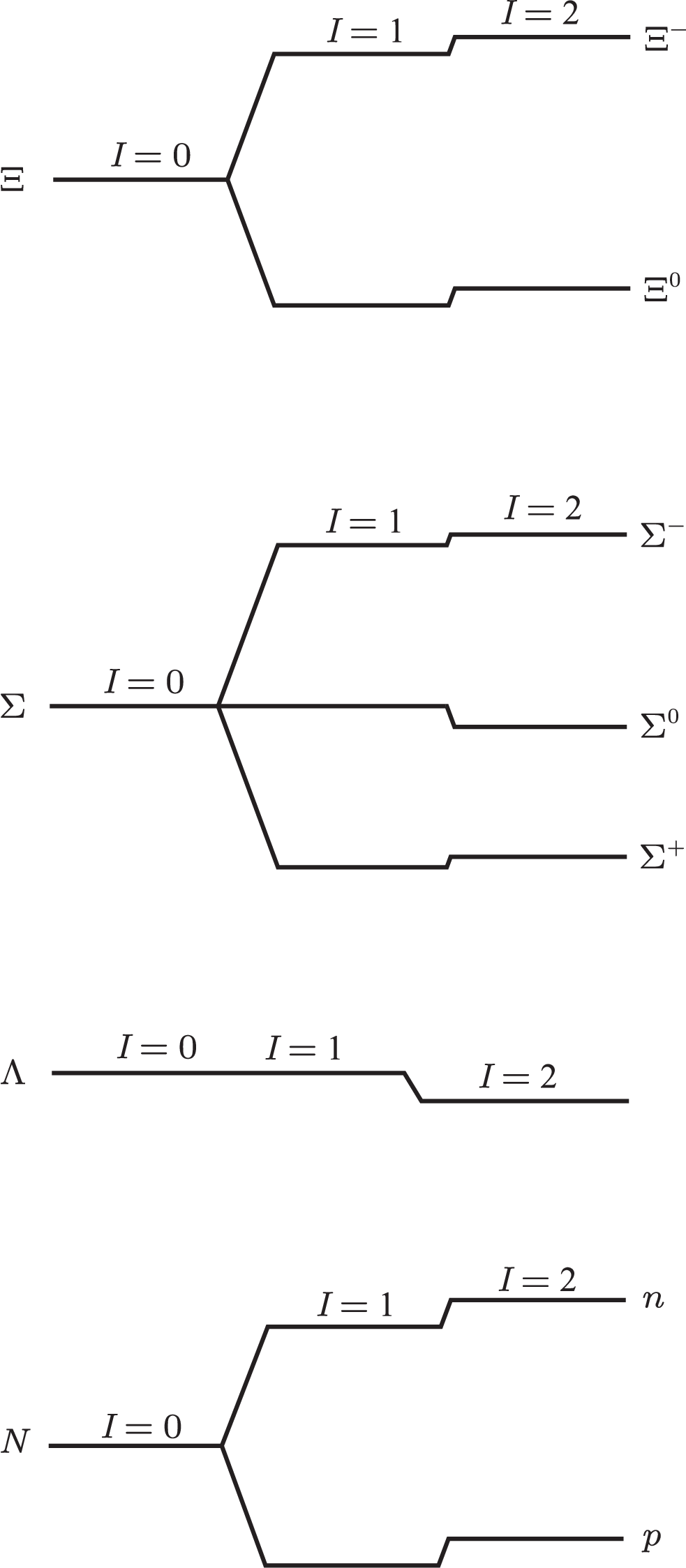}}
\caption{\label{fig:mh1}Schematic representation of mass splittings in the different channels $I=0$, $1$, $2$ of baryon octet.}
\end{figure}

\begin{figure}[ht]
\scalebox{0.30}{\includegraphics{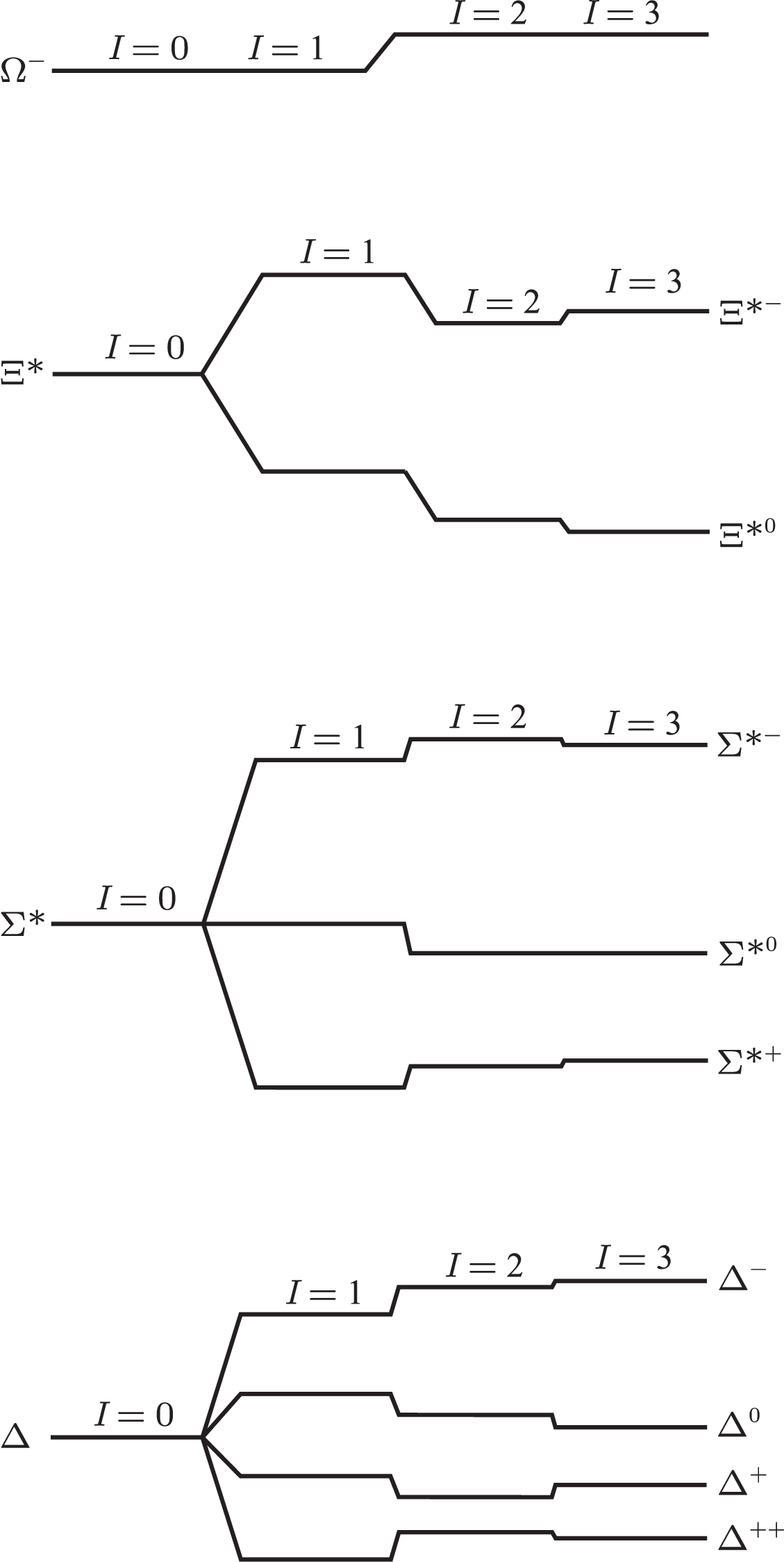}}
\caption{\label{fig:mh2}Schematic representation of mass splittings in the different channels $I=0$, $1$, $2$, $3$ of baryon decuplet.}
\end{figure}

To close this section, it can be concluded that the $1/N_c$ expansion provides a strong evidence for a mass hierarchy in baryons, as it was pointed out in Ref.~\cite{jl}. This can be better appreciated in the schematic representations of mass splittings displayed in Figs.!\ref{fig:mh1} and \ref{fig:mh2} for baryon octet and baryon decuplet, respectively.

\section{\label{sec:con}Concluding remarks}

The lowest-lying baryon masses are studied by considering a combined expansion in $1/N_c$ and perturbative symmetry breaking corrections. Two main sources account for the latter, namely, the quark mass differences and electromagnetic contributions. Thus, from a group theory point of view, multiple tensor products of $SU(3)$ adjoint representations are required to incorporate all these effects in a systematic and organized way. This goal is achieved by constructing the most general $1/N_c$ expansions with up to three free flavor indices. Appropriate setting of these free indices yields the different $I=0$, $1$, $2$ and $3$ channels considered in the analysis. The calculation is facilitated by the use of flavor projection operators \cite{banda1,banda2}, which allow one to separate all irreducible representations involved in SB, namely, $1$, $8$, $10+\overline{10}$, $27$, $35+\overline{35}$ and $64$. An additional simplification is achieved by observing that practically all operator coefficients that appear in the analysis can be absorbed in terms of a few effective ones, 19 in total.

The use of experimental \cite{part} and numerical \cite{romiti} data allows one to perform a least-squares fit to explore these effective parameters. The fit yields the values listed in (\ref{eq:bestfit}). Although the best-fit values are roughly in accord to expectations from the $1/N_c$ expansion itself, namely, the $1$ representation is the least-suppressed contribution, followed by the $8$, $10+\overline{10}$, $27$, and $64$, the fit somehow is not entirely satisfactory so it can not be regarded as definitive. The values of $\Delta$ mass differences from Ref.~\cite{romiti}, along with the measured masses recommended in Ref.~\cite{part} are the data used in the fit. Thus, the $\Delta$ masses listed in Table~\ref{t:mass} are actual predictions of the approach, which are in accord to the ones presented in that reference. Another issue is constituted by the poorly determined operator coefficients that come from the $64$ representation.

On general grounds, the mass determinations obtained in the $1/N_c$ expansion including SB corrections give robustness to the mass hierarchy in the baryon sector pointed out in Ref.~\cite{jl}. The schematic representation of mass splitting of Figs.~\ref{fig:mh1} and \ref{fig:mh2} allows one to better appreciate this fact.

To close this paper, it should be mention that a common procedure advocated in the literature is to separate electromagnetic and strong isospin breaking effects to give two separate values. However, the separation may induce an ambiguity due to the mechanism by which they are separated; in a few words, a model-dependence may be induced. In the present analysis, these two effects are present in the baryon mass expansion and can not be untangled a priori. By no means this can be regarded as a failure of the approach.

\begin{acknowledgments}
The authors are grateful to Consejo Nacional de Ciencia y Tecnolog{\'\i}a (Mexico) for support through the {\it Ciencia de Frontera} project CF-2023-I-162.

\end{acknowledgments}

\appendix

\section{\label{app:q1q2q3}Analytic expressions for projection operators acting on three adjoints}

The flavor projection operators $[\tilde{\mathcal{P}}^{(m)}]^{a_1a_2a_3b_1b_2b_3}$, (\ref{eq:pa_1a_2a_3}), when acting on the product of three adjoints $Q_1^{b_1}Q_2^{b_2}Q_3^{b_3}$, effectively project out that piece of $Q_1^{a_1}Q_2^{a_2}Q_3^{a_3}$ transforming in the irreducible representation of dimension $m$ of $SU(3)$, according to decomposition (\ref{eq:8x8x8}).

The specific structures required by the analysis of baryon mass splittings, $I = 0,1,2,3$, for which $\{a_1,a_2,a_3\} = \{8,8,8\}$, $\{a_1,a_2,a_3\} = \{3,8,8\}$, $\{a_1,a_2,a_3\} = \{3,3,8\}$, and $\{a_1,a_2,a_3\} = \{3,3,3\}$, respectively, are listed below.

\subsection{$I = 0$}

\begin{eqnarray}
[\tilde{\mathcal{P}}^{(1)} Q_1 Q_2 Q_3]^{888} & = & - \frac{1}{40} Q_1^1 Q_2^1 Q_3^8 - \frac{\sqrt{3}}{80} Q_1^1 Q_2^4 Q_3^6 - \frac{\sqrt{3}}{80} Q_1^1 Q_2^5 Q_3^7 - \frac{\sqrt{3}}{80} Q_1^1 Q_2^6 Q_3^4 - \frac{\sqrt{3}}{80} Q_1^1 Q_2^7 Q_3^5 \nonumber \\
& & \mbox{} - \frac{1}{40} Q_1^1 Q_2^8 Q_3^1 - \frac{1}{40} Q_1^2 Q_2^2 Q_3^8 + \frac{\sqrt{3}}{80} Q_1^2 Q_2^4 Q_3^7 - \frac{\sqrt{3}}{80} Q_1^2 Q_2^5 Q_3^6 - \frac{\sqrt{3}}{80} Q_1^2 Q_2^6 Q_3^5 \nonumber \\
& & \mbox{} + \frac{\sqrt{3}}{80} Q_1^2 Q_2^7 Q_3^4 - \frac{1}{40} Q_1^2 Q_2^8 Q_3^2 - \frac{1}{40} Q_1^3 Q_2^3 Q_3^8 - \frac{\sqrt{3}}{80} Q_1^3 Q_2^4 Q_3^4 - \frac{\sqrt{3}}{80} Q_1^3 Q_2^5 Q_3^5 \nonumber \\
& & \mbox{} + \frac{\sqrt{3}}{80} Q_1^3 Q_2^6 Q_3^6 + \frac{\sqrt{3}}{80} Q_1^3 Q_2^7 Q_3^7 - \frac{1}{40} Q_1^3 Q_2^8 - \frac{\sqrt{3}}{80} Q_1^4 Q_2^1 Q_3^6 + \frac{\sqrt{3}}{80} Q_1^4 Q_2^2 Q_3^7 \nonumber \\
& & \mbox{} - \frac{\sqrt{3}}{80} Q_1^4 Q_2^3 Q_3^4 - \frac{\sqrt{3}}{80} Q_1^4 Q_2^4 Q_3^3 + \frac{1}{80} Q_1^4 Q_2^4 Q_3^8 - \frac{\sqrt{3}}{80} Q_1^4 Q_2^6 Q_3^1 + \frac{\sqrt{3}}{80} Q_1^4 Q_2^7 Q_3^2 \nonumber \\
& & \mbox{} + \frac{1}{80} Q_1^4 Q_2^8 Q_3^4 - \frac{\sqrt{3}}{80} Q_1^5 Q_2^1 Q_3^7 - \frac{\sqrt{3}}{80} Q_1^5 Q_2^2 Q_3^6 - \frac{\sqrt{3}}{80} Q_1^5 Q_2^3 Q_3^5 - \frac{\sqrt{3}}{80} Q_1^5 Q_2^5 Q_3^3 \nonumber \\
& & \mbox{} + \frac{1}{80} Q_1^5 Q_2^5 Q_3^8 - \frac{\sqrt{3}}{80} Q_1^5 Q_2^6 Q_3^2 - \frac{\sqrt{3}}{80} Q_1^5 Q_2^7 Q_3^1 + \frac{1}{80} Q_1^5 Q_2^8 Q_3^5 - \frac{\sqrt{3}}{80} Q_1^6 Q_2^1 Q_3^4 \nonumber \\
& & \mbox{} - \frac{\sqrt{3}}{80} Q_1^6 Q_2^2 Q_3^5 + \frac{\sqrt{3}}{80} Q_1^6 Q_2^3 Q_3^6 - \frac{\sqrt{3}}{80} Q_1^6 Q_2^4 Q_3^1 - \frac{\sqrt{3}}{80} Q_1^6 Q_2^5 Q_3^2 + \frac{\sqrt{3}}{80} Q_1^6 Q_2^6 Q_3^3 \nonumber \\
& & \mbox{} + \frac{1}{80} Q_1^6 Q_2^6 Q_3^8 + \frac{1}{80} Q_1^6 Q_2^8 Q_3^6 - \frac{\sqrt{3}}{80} Q_1^7 Q_2^1 Q_3^5 + \frac{\sqrt{3}}{80} Q_1^7 Q_2^2 Q_3^4 + \frac{\sqrt{3}}{80} Q_1^7 Q_2^3 Q_3^7 \nonumber \\
& & \mbox{} + \frac{\sqrt{3}}{80} Q_1^7 Q_2^4 Q_3^2 - \frac{\sqrt{3}}{80} Q_1^7 Q_2^5 Q_3^1 + \frac{\sqrt{3}}{80}
 Q_1^7 Q_2^7 Q_3^3 + \frac{1}{80} Q_1^7 Q_2^7 Q_3^8 + \frac{1}{80} Q_1^7 Q_2^8 Q_3^7 \nonumber \\
& & \mbox{} - \frac{1}{40} Q_1^8 Q_2^1 Q_3^1 - \frac{1}{40} Q_1^8 Q_2^2 Q_3^2 - \frac{1}{40} Q_1^8 Q_2^3 Q_3^3 + \frac{1}{80} Q_1^8 Q_2^4 Q_3^4 + \frac{1}{80} Q_1^8 Q_2^5 Q_3^5 \nonumber \\
& & \mbox{} + \frac{1}{80} Q_1^8 Q_2^6 Q_3^6 + \frac{1}{80} Q_1^8 Q_2^7 Q_3^7 + \frac{1}{40} Q_1^8 Q_2^8 Q_3^8,
\end{eqnarray}

\begin{eqnarray}
[\tilde{\mathcal{P}}^{(8)} Q_1 Q_2 Q_3]^{888} & = & \frac{1}{10} Q_1^1 Q_2^1 Q_3^8 + \frac{1}{10} Q_1^1 Q_2^8 Q_3^1 + \frac{1}{10} Q_1^2 Q_2^2 Q_3^8 + \frac{1}{10} Q_1^2 Q_2^8 Q_3^2 + \frac{1}{10} Q_1^3 Q_2^3 Q_3^8 \nonumber \\
& & \mbox{} + \frac{1}{10} Q_1^3 Q_2^8 Q_3^3 + \frac{1}{10} Q_1^4 Q_2^4 Q_3^8 + \frac{1}{10} Q_1^4 Q_2^8 Q_3^4 + \frac{1}{10} Q_1^5 Q_2^5 Q_3^8 + \frac{1}{10} Q_1^5 Q_2^8 Q_3^5 \nonumber \\
& & \mbox{} + \frac{1}{10} Q_1^6 Q_2^6 Q_3^8 + \frac{1}{10} Q_1^6 Q_2^8 Q_3^6 + \frac{1}{10} Q_1^7 Q_2^7 Q_3^8 + \frac{1}{10} Q_1^7 Q_2^8 Q_3^7 + \frac{1}{10} Q_1^8 Q_2^1 Q_3^1 \nonumber \\
& & \mbox{} + \frac{1}{10} Q_1^8 Q_2^2 Q_3^2 + \frac{1}{10} Q_1^8 Q_2^3 Q_3^3 + \frac{1}{10} Q_1^8 Q_2^4 Q_3^4 + \frac{1}{10} Q_1^8 Q_2^5 Q_3^5 + \frac{1}{10} Q_1^8 Q_2^6 Q_3^6 \nonumber \\
& & \mbox{} + \frac{1}{10} Q_1^8 Q_2^7 Q_3^7 + \frac{3}{10} Q_1^8 Q_2^8 Q_3^8,
\end{eqnarray}

\begin{equation}
[\tilde{\mathcal{P}}^{(10+\overline{10})} Q_1 Q_2 Q_3]^{888} = 0,
\end{equation}

\begin{eqnarray}
[\tilde{\mathcal{P}}^{(27)} Q_1 Q_2 Q_3]^{888} & = & - \frac{33}{280} Q_1^1 Q_2^1 Q_3^8 + \frac{3\sqrt{3}}{112} Q_1^1 Q_2^4 Q_3^6 + \frac{3\sqrt{3}}{112} Q_1^1 Q_2^5 Q_3^7 + \frac{3\sqrt{3}}{112} Q_1^1 Q_2^6 Q_3^4 + \frac{3\sqrt{3}}{112} Q_1^1 Q_2^7 Q_3^5 \nonumber \\
& & \mbox{} - \frac{33}{280} Q_1^1 Q_2^8 Q_3^1 - \frac{33}{280} Q_1^2 Q_2^2 Q_3^8 - \frac{3\sqrt{3}}{112} Q_1^2 Q_2^4 Q_3^7 + \frac{3\sqrt{3}}{112} Q_1^2 Q_2^5 Q_3^6 + \frac{3\sqrt{3}}{112} Q_1^2 Q_2^6 Q_3^5 \nonumber \\
& & \mbox{} - \frac{3\sqrt{3}}{112} Q_1^2 Q_2^7 Q_3^4 - \frac{33}{280} Q_1^2 Q_2^8 Q_3^2 - \frac{33}{280} Q_1^3 Q_2^3 Q_3^8 + \frac{3\sqrt{3}}{112} Q_1^3 Q_2^4 Q_3^4 + \frac{3\sqrt{3}}{112} Q_1^3 Q_2^5 Q_3^5 \nonumber \\
& & \mbox{} - \frac{3\sqrt{3}}{112} Q_1^3 Q_2^6 Q_3^6 - \frac{3\sqrt{3}}{112} Q_1^3 Q_2^7 Q_3^7 - \frac{33}{280} Q_1^3 Q_2^8 Q_3^3 + \frac{3\sqrt{3}}{112} Q_1^4 Q_2^1 Q_3^6 - \frac{3\sqrt{3}}{112} Q_1^4 Q_2^2 Q_3^7 \nonumber \\
& & \mbox{} + \frac{3\sqrt{3}}{112} Q_1^4 Q_2^3 Q_3^4 + \frac{3\sqrt{3}}{112} Q_1^4 Q_2^4 Q_3^3 + \frac{9}{560} Q_1^4 Q_2^4 Q_3^8 + \frac{3\sqrt{3}}{112} Q_1^4 Q_2^6 Q_3^1 - \frac{3\sqrt{3}}{112} Q_1^4 Q_2^7 Q_3^2 \nonumber \\
& & \mbox{} + \frac{9}{560} Q_1^4 Q_2^8 Q_3^4 + \frac{3\sqrt{3}}{112} Q_1^5 Q_2^1 Q_3^7 + \frac{3\sqrt{3}}{112} Q_1^5 Q_2^2 Q_3^6 + \frac{3\sqrt{3}}{112} Q_1^5 Q_2^3 Q_3^5 + \frac{3\sqrt{3}}{112} Q_1^5 Q_2^5 Q_3^3 \nonumber \\
& & \mbox{} + \frac{9}{560} Q_1^5 Q_2^5 Q_3^8 + \frac{3\sqrt{3}}{112} Q_1^5 Q_2^6 Q_3^2 + \frac{3\sqrt{3}}{112} Q_1^5 Q_2^7 Q_3^1 + \frac{9}{560} Q_1^5 Q_2^8 Q_3^5 + \frac{3\sqrt{3}}{112} Q_1^6 Q_2^1 Q_3^4 \nonumber \\
& & \mbox{} + \frac{3\sqrt{3}}{112} Q_1^6 Q_2^2 Q_3^5 - \frac{3\sqrt{3}}{112} Q_1^6 Q_2^3 Q_3^6 + \frac{3\sqrt{3}}{112} Q_1^6 Q_2^4 Q_3^1 + \frac{3\sqrt{3}}{112} Q_1^6 Q_2^5 Q_3^2 - \frac{3\sqrt{3}}{112} Q_1^6 Q_2^6 Q_3^3 \nonumber \\
& & \mbox{} + \frac{9}{560} Q_1^6 Q_2^6 Q_3^8 + \frac{9}{560} Q_1^6 Q_2^8 Q_3^6 + \frac{3\sqrt{3}}{112} Q_1^7 Q_2^1 Q_3^5 - \frac{3\sqrt{3}}{112} Q_1^7 Q_2^2 Q_3^4 - \frac{3\sqrt{3}}{112} Q_1^7 Q_2^3 Q_3^7 \nonumber \\
& & \mbox{} - \frac{3\sqrt{3}}{112} Q_1^7 Q_2^4 Q_3^2 + \frac{3\sqrt{3}}{112} Q_1^7 Q_2^5 Q_3^1 - \frac{3\sqrt{3}}{112} Q_1^7 Q_2^7 Q_3^3 + \frac{9}{560} Q_1^7 Q_2^7 Q_3^8 + \frac{9}{560} Q_1^7 Q_2^8 Q_3^7 \nonumber \\
& & \mbox{} - \frac{33}{280} Q_1^8 Q_2^1 Q_3^1 - \frac{33}{280} Q_1^8 Q_2^2 Q_3^2 - \frac{33}{280} Q_1^8 Q_2^3 Q_3^3 + \frac{9}{560} Q_1^8 Q_2^4 Q_3^4 + \frac{9}{560} Q_1^8 Q_2^5 Q_3^5 \nonumber \\
& & \mbox{} + \frac{9}{560} Q_1^8 Q_2^6 Q_3^6 + \frac{9}{560} Q_1^8 Q_2^7 Q_3^7 + \frac{81}{280} Q_1^8 Q_2^8 Q_3^8,
\end{eqnarray}

\begin{equation}
[\tilde{\mathcal{P}}^{(35+\overline{35})} Q_1 Q_2 Q_3]^{888} = 0,
\end{equation}

\begin{eqnarray}
[\tilde{\mathcal{P}}^{(64)} Q_1 Q_2 Q_3]^{888} & = & \frac{3}{70} Q_1^1 Q_2^1 Q_3^8 - \frac{\sqrt{3}}{70} Q_1^1 Q_2^4 Q_3^6 - \frac{\sqrt{3}}{70} Q_1^1 Q_2^5 Q_3^7 - \frac{\sqrt{3}}{70} Q_1^1 Q_2^6 Q_3^4 - \frac{\sqrt{3}}{70} Q_1^1 Q_2^7 Q_3^5 \nonumber \\
& & \mbox{} + \frac{3}{70} Q_1^1 Q_2^8 Q_3^1 + \frac{3}{70} Q_1^2 Q_2^2 Q_3^8 + \frac{\sqrt{3}}{70} Q_1^2 Q_2^4 Q_3^7 - \frac{\sqrt{3}}{70} Q_1^2 Q_2^5 Q_3^6 - \frac{\sqrt{3}}{70} Q_1^2 Q_2^6 Q_3^5 \nonumber \\
& & \mbox{} + \frac{\sqrt{3}}{70} Q_1^2 Q_2^7 Q_3^4 + \frac{3}{70} Q_1^2 Q_2^8 Q_3^2 + \frac{3}{70} Q_1^3 Q_2^3 Q_3^8 - \frac{\sqrt{3}}{70} Q_1^3 Q_2^4 Q_3^4 - \frac{\sqrt{3}}{70} Q_1^3 Q_2^5 Q_3^5 \nonumber \\
& & \mbox{} + \frac{\sqrt{3}}{70} Q_1^3 Q_2^6 Q_3^6 + \frac{\sqrt{3}}{70} Q_1^3 Q_2^7 Q_3^7 + \frac{3}{70} Q_1^3 Q_2^8 Q_3^3 - \frac{\sqrt{3}}{70} Q_1^4 Q_2^1 Q_3^6 + \frac{\sqrt{3}}{70} Q_1^4 Q_2^2 Q_3^7 \nonumber \\
& & \mbox{} - \frac{\sqrt{3}}{70} Q_1^4 Q_2^3 Q_3^4 - \frac{\sqrt{3}}{70} Q_1^4 Q_2^4 Q_3^3 - \frac{9}{70} Q_1^4 Q_2^4 Q_3^8 - \frac{\sqrt{3}}{70} Q_1^4 Q_2^6 Q_3^1 +\frac{\sqrt{3}}{70} Q_1^4 Q_2^7 Q_3^2 \nonumber \\
& & \mbox{} - \frac{9}{70} Q_1^4 Q_2^8 Q_3^4 - \frac{\sqrt{3}}{70} Q_1^5 Q_2^1 Q_3^7 - \frac{\sqrt{3}}{70} Q_1^5 Q_2^2 Q_3^6 - \frac{\sqrt{3}}{70} Q_1^5 Q_2^3 Q_3^5 - \frac{\sqrt{3}}{70} Q_1^5 Q_2^5 Q_3^3 \nonumber \\
& & \mbox{} - \frac{9}{70} Q_1^5 Q_2^5 Q_3^8 - \frac{\sqrt{3}}{70} Q_1^5 Q_2^6 Q_3^2 - \frac{\sqrt{3}}{70} Q_1^5 Q_2^7 Q_3^1 - \frac{9}{70} Q_1^5 Q_2^8 Q_3^5 - \frac{\sqrt{3}}{70} Q_1^6 Q_2^1 Q_3^4 \nonumber \\
& & \mbox{} - \frac{\sqrt{3}}{70} Q_1^6 Q_2^2 Q_3^5 + \frac{\sqrt{3}}{70} Q_1^6 Q_2^3 Q_3^6 - \frac{\sqrt{3}}{70} Q_1^6 Q_2^4 Q_3^1 - \frac{\sqrt{3}}{70} Q_1^6 Q_2^5 Q_3^2 + \frac{\sqrt{3}}{70} Q_1^6 Q_2^6 Q_3^3 \nonumber \\
& & \mbox{} - \frac{9}{70} Q_1^6 Q_2^6 Q_3^8 - \frac{9}{70} Q_1^6 Q_2^8 Q_3^6 - \frac{\sqrt{3}}{70} Q_1^7 Q_2^1 Q_3^5 + \frac{\sqrt{3}}{70} Q_1^7 Q_2^2 Q_3^4 + \frac{\sqrt{3}}{70} Q_1^7 Q_2^3 Q_3^7 \nonumber \\
& & \mbox{} + \frac{\sqrt{3}}{70} Q_1^7 Q_2^4 Q_3^2 - \frac{\sqrt{3}}{70} Q_1^7 Q_2^5 Q_3^1 + \frac{\sqrt{3}}{70} Q_1^7 Q_2^7 Q_3^3 - \frac{9}{70} Q_1^7 Q_2^7 Q_3^8 - \frac{9}{70} Q_1^7 Q_2^8 Q_3^7 \nonumber \\
& & \mbox{} + \frac{3}{70} Q_1^8 Q_2^1 Q_3^1 + \frac{3}{70} Q_1^8 Q_2^2 Q_3^2 + \frac{3}{70} Q_1^8 Q_2^3 Q_3^3 - \frac{9}{70} Q_1^8 Q_2^4 Q_3^4 - \frac{9}{70} Q_1^8 Q_2^5 Q_3^5 \nonumber \\
& & \mbox{} - \frac{9}{70} Q_1^8 Q_2^6 Q_3^6 - \frac{9}{70} Q_1^8 Q_2^7 Q_3^7 + \frac{27}{70} Q_1^8 Q_2^8 Q_3^8.
\end{eqnarray}

\subsection{$I = 1$}

\begin{equation}
[\tilde{\mathcal{P}}^{(1)} Q_1 Q_2 Q_3]^{388} = 0,
\end{equation}

\begin{eqnarray}
[\tilde{\mathcal{P}}^{(8)} Q_1 Q_2 Q_3]^{388} & = & \frac{1}{90} Q_1^1 Q_2^1 Q_3^3 + \frac{1}{90} Q_1^1 Q_2^3 Q_3^1 + \frac{1}{90} Q_1^2 Q_2^2 Q_3^3 + \frac{1}{90} Q_1^2 Q_2^3 Q_3^2 + \frac{7}{90} Q_1^3 Q_2^1 Q_3^1 \nonumber \\
& & \mbox{} + \frac{7}{90} Q_1^3 Q_2^2 Q_3^2 + \frac{1}{10} Q_1^3 Q_2^3 Q_3^3 + \frac{13}{90} Q_1^3 Q_2^4 Q_3^4 + \frac{13}{90} Q_1^3 Q_2^5 Q_3^5 + \frac{13}{90} Q_1^3 Q_2^6 Q_3^6 \nonumber \\
& & \mbox{} + \frac{13}{90} Q_1^3 Q_2^7 Q_3^7 + \frac16 Q_1^3 Q_2^8 Q_3^8 - \frac{1}{30} Q_1^4 Q_2^1 Q_3^6 + \frac{1}{30} Q_1^4 Q_2^2 Q_3^7 - \frac{1}{45} Q_1^4 Q_2^3 Q_3^4 \nonumber \\
& & \mbox{} - \frac{1}{45} Q_1^4 Q_2^4 Q_3^3 + \frac{\sqrt{3}}{90} Q_1^4 Q_2^4 Q_3^8 - \frac{1}{30} Q_1^4 Q_2^6 Q_3^1 + \frac{1}{30} Q_1^4 Q_2^7 Q_3^2 + \frac{\sqrt{3}}{90} Q_1^4 Q_2^8 Q_3^4 \nonumber \\
& & \mbox{} - \frac{1}{30} Q_1^5 Q_2^1 Q_3^7 - \frac{1}{30} Q_1^5 Q_2^2 Q_3^6 - \frac{1}{45} Q_1^5 Q_2^3 Q_3^5 - \frac{1}{45} Q_1^5 Q_2^5 Q_3^3 + \frac{\sqrt{3}}{90} Q_1^5 Q_2^5 Q_3^8 \nonumber \\
& & \mbox{} - \frac{1}{30} Q_1^5 Q_2^6 Q_3^2 - \frac{1}{30} Q_1^5 Q_2^7 Q_3^1 + \frac{\sqrt{3}}{90} Q_1^5 Q_2^8 Q_3^5 + \frac{1}{30} Q_1^6 Q_2^1 Q_3^4 + \frac{1}{30} Q_1^6 Q_2^2 Q_3^5 \nonumber \\
& & \mbox{} - \frac{1}{45} Q_1^6 Q_2^3 Q_3^6 + \frac{1}{30} Q_1^6 Q_2^4 Q_3^1 + \frac{1}{30} Q_1^6 Q_2^5 Q_3^2 - \frac{1}{45} Q_1^6 Q_2^6 Q_3^3 - \frac{\sqrt{3}}{90} Q_1^6 Q_2^6 Q_3^8 \nonumber \\
& & \mbox{} - \frac{\sqrt{3}}{90} Q_1^6 Q_2^8 Q_3^6 + \frac{1}{30} Q_1^7 Q_2^1 Q_3^5 - \frac{1}{30} Q_1^7 Q_2^2 Q_3^4 - \frac{1}{45} Q_1^7 Q_2^3 Q_3^7 - \frac{1}{30} Q_1^7 Q_2^4 Q_3^2 \nonumber \\
& & \mbox{} + \frac{1}{30} Q_1^7 Q_2^5 Q_3^1 - \frac{1}{45} Q_1^7 Q_2^7 Q_3^3 - \frac{\sqrt{3}}{90} Q_1^7 Q_2^7 Q_3^8 - \frac{\sqrt{3}}{90} Q_1^7 Q_2^8 Q_3^7 - \frac{1}{30} Q_1^8 Q_2^3 Q_3^8 \nonumber \\
& & \mbox{} - \frac{\sqrt{3}}{45} Q_1^8 Q_2^4 Q_3^4 - \frac{\sqrt{3}}{45} Q_1^8 Q_2^5 Q_3^5 + \frac{\sqrt{3}}{45} Q_1^8 Q_2^6 Q_3^6 + \frac{\sqrt{3}}{45} Q_1^8 Q_2^7 Q_3^7 - \frac{1}{30} Q_1^8 Q_2^8 Q_3^3,
\end{eqnarray}

\begin{eqnarray}
[\tilde{\mathcal{P}}^{(10+\overline{10})} Q_1 Q_2 Q_3]^{388} & = & - \frac{1}{18} Q_1^1 Q_2^1 Q_3^3 - \frac{1}{18} Q_1^1 Q_2^3 Q_3^1 - \frac{1}{18} Q_1^2 Q_2^2 Q_3^3 - \frac{1}{18} Q_1^2 Q_2^3 Q_3^2 - \frac{1}{18} Q_1^3 Q_2^1 Q_3^1 \nonumber \\
& & \mbox{} - \frac{1}{18} Q_1^3 Q_2^2 Q_3^2 - \frac16 Q_1^3 Q_2^3 Q_3^3 + \frac{1}{36} Q_1^3 Q_2^4 Q_3^4 + \frac{1}{36} Q_1^3 Q_2^5 Q_3^5 + \frac{1}{36} Q_1^3 Q_2^6 Q_3^6 \nonumber \\
& & \mbox{} + \frac{1}{36} Q_1^3 Q_2^7 Q_3^7 + \frac16 Q_1^3 Q_2^8 Q_3^8 + \frac{1}{36} Q_1^4 Q_2^3 Q_3^4 + \frac{1}{36} Q_1^4 Q_2^4 Q_3^3 + \frac{\sqrt{3}}{36} Q_1^4 Q_2^4 Q_3^8 \nonumber \\
& & \mbox{} + \frac{\sqrt{3}}{36} Q_1^4 Q_2^8 Q_3^4 + \frac{1}{36} Q_1^5 Q_2^3 Q_3^5 + \frac{1}{36} Q_1^5 Q_2^5 Q_3^3 + \frac{\sqrt{3}}{36} Q_1^5 Q_2^5 Q_3^8 + \frac{\sqrt{3}}{36} Q_1^5 Q_2^8 Q_3^5 \nonumber \\
& & \mbox{} + \frac{1}{36} Q_1^6 Q_2^3 Q_3^6 + \frac{1}{36} Q_1^6 Q_2^6 Q_3^3 - \frac{\sqrt{3}}{36} Q_1^6 Q_2^6 Q_3^8 - \frac{\sqrt{3}}{36} Q_1^6 Q_2^8 Q_3^6 + \frac{1}{36} Q_1^7 Q_2^3 Q_3^7 \nonumber \\
& & \mbox{} + \frac{1}{36} Q_1^7 Q_2^7 Q_3^3 - \frac{\sqrt{3}}{36} Q_1^7 Q_2^7 Q_3^8 - \frac{\sqrt{3}}{36} Q_1^7 Q_2^8 Q_3^7 + \frac16 Q_1^8 Q_2^3 Q_3^8 + \frac{\sqrt{3}}{36} Q_1^8 Q_2^4 Q_3^4 \nonumber \\
& & \mbox{} + \frac{\sqrt{3}}{36} Q_1^8 Q_2^5 Q_3^5 - \frac{\sqrt{3}}{36} Q_1^8 Q_2^6 Q_3^6 - \frac{\sqrt{3}}{36} Q_1^8 Q_2^7 Q_3^7 + \frac16 Q_1^8 Q_2^8 Q_3^3,
\end{eqnarray}

\begin{eqnarray}
[\tilde{\mathcal{P}}^{(27)} Q_1 Q_2 Q_3]^{388} & = & \frac{9}{140} Q_1^1 Q_2^1 Q_3^3 + \frac{9}{140} Q_1^1 Q_2^3 Q_3^1 + \frac{9}{140}
 Q_1^2 Q_2^2 Q_3^3 + \frac{9}{140} Q_1^2 Q_2^3 Q_3^2 - \frac{3}{35} Q_1^3 Q_2^1 Q_3^1 \nonumber \\
& & \mbox{} - \frac{3}{35} Q_1^3 Q_2^2 Q_3^2 + \frac{3}{70} Q_1^3 Q_2^3 Q_3^3 - \frac{3}{140} Q_1^3 Q_2^4 Q_3^4 - \frac{3}{140} Q_1^3 Q_2^5 Q_3^5 - \frac{3}{140} Q_1^3 Q_2^6 Q_3^6 \nonumber \\
& & \mbox{} - \frac{3}{140} Q_1^3 Q_2^7 Q_3^7 + \frac{3}{14} Q_1^3 Q_2^8 Q_3^8 + \frac{3}{40} Q_1^4 Q_2^1 Q_3^6 - \frac{3}{40} Q_1^4 Q_2^2 Q_3^7 - \frac{3}{140} Q_1^4 Q_2^3 Q_3^4 \nonumber \\
& & \mbox{} - \frac{3}{140} Q_1^4 Q_2^4 Q_3^3 + \frac{\sqrt{3}}{35} Q_1^4 Q_2^4 Q_3^8 + \frac{3}{40} Q_1^4 Q_2^6 Q_3^1 - \frac{3}{40} Q_1^4 Q_2^7 Q_3^2 + \frac{\sqrt{3}}{35} Q_1^4 Q_2^8 Q_3^4 \nonumber \\
& & \mbox{} + \frac{3}{40} Q_1^5 Q_2^1 Q_3^7 + \frac{3}{40} Q_1^5 Q_2^2 Q_3^6 - \frac{3}{140} Q_1^5 Q_2^3 Q_3^5 - \frac{3}{140} Q_1^5 Q_2^5 Q_3^3 + \frac{\sqrt{3}}{35} Q_1^5 Q_2^5 Q_3^8 \nonumber \\
& & \mbox{} + \frac{3}{40} Q_1^5 Q_2^6 Q_3^2 + \frac{3}{40} Q_1^5 Q_2^7 Q_3^1 + \frac{\sqrt{3}}{35} Q_1^5 Q_2^8 Q_3^5 - \frac{3}{40} Q_1^6 Q_2^1 Q_3^4 - \frac{3}{40} Q_1^6 Q_2^2 Q_3^5 \nonumber \\
& & \mbox{} - \frac{3}{140} Q_1^6 Q_2^3 Q_3^6 - \frac{3}{40} Q_1^6 Q_2^4 Q_3^1 - \frac{3}{40} Q_1^6 Q_2^5 Q_3^2 - \frac{3}{140} Q_1^6 Q_2^6 Q_3^3 - \frac{\sqrt{3}}{35} Q_1^6 Q_2^6 Q_3^8 \nonumber \\
& & \mbox{} - \frac{\sqrt{3}}{35} Q_1^6 Q_2^8 Q_3^6 - \frac{3}{40} Q_1^7 Q_2^1 Q_3^5 + \frac{3}{40} Q_1^7 Q_2^2 Q_3^4 - \frac{3}{140}
 Q_1^7 Q_2^3 Q_3^7 + \frac{3}{40} Q_1^7 Q_2^4 Q_3^2 \nonumber \\
& & \mbox{} - \frac{3}{40} Q_1^7 Q_2^5 Q_3^1 - \frac{3}{140} Q_1^7 Q_2^7 Q_3^3 - \frac{\sqrt{3}}{35} Q_1^7 Q_2^7 Q_3^8 - \frac{\sqrt{3}}{35} Q_1^7 Q_2^8 Q_3^7 - \frac{3}{35} Q_1^8 Q_2^3 Q_3^8 \nonumber \\
& & \mbox{} - \frac{3\sqrt{3}}{140} Q_1^8 Q_2^4 Q_3^4 - \frac{3\sqrt{3}}{140} Q_1^8 Q_2^5 Q_3^5 + \frac{3\sqrt{3}}{140} Q_1^8 Q_2^6 Q_3^6 + \frac{3\sqrt{3}}{140} Q_1^8 Q_2^7 Q_3^7 - \frac{3}{35} Q_1^8 Q_2^8 Q_3^3, \nonumber \\
\end{eqnarray}

\begin{eqnarray}
[\tilde{\mathcal{P}}^{(35+\overline{35})} Q_1 Q_2 Q_3]^{388} & = & - \frac{1}{36} Q_1^1 Q_2^1 Q_3^3 - \frac{1}{36} Q_1^1 Q_2^3 Q_3^1 - \frac{1}{36} Q_1^2 Q_2^2 Q_3^3 - \frac{1}{36} Q_1^2 Q_2^3 Q_3^2 + \frac{1}{18} Q_1^3 Q_2^1 Q_3^1 \nonumber \\
& & \mbox{} + \frac{1}{18} Q_1^3 Q_2^2 Q_3^2 - \frac19 Q_1^3 Q_2^4 Q_3^4 - \frac19 Q_1^3 Q_2^5 Q_3^5 - \frac19 Q_1^3 Q_2^6 Q_3^6 - \frac19 Q_1^3 Q_2^7 Q_3^7 \nonumber \\
& & \mbox{} + \frac13 Q_1^3 Q_2^8 Q_3^8 - \frac{1}{24} Q_1^4 Q_2^1 Q_3^6 + \frac{1}{24} Q_1^4 Q_2^2 Q_3^7 + \frac{1}{18} Q_1^4 Q_2^3 Q_3^4 + \frac{1}{18} Q_1^4 Q_2^4 Q_3^3 \nonumber \\
& & \mbox{} - \frac{\sqrt{3}}{36} Q_1^4 Q_2^4 Q_3^8 - \frac{1}{24} Q_1^4 Q_2^6 Q_3^1 + \frac{1}{24} Q_1^4 Q_2^7 Q_3^2 - \frac{\sqrt{3}}{36} Q_1^4 Q_2^8 Q_3^4 - \frac{1}{24} Q_1^5 Q_2^1 Q_3^7 \nonumber \\
& & \mbox{} - \frac{1}{24} Q_1^5 Q_2^2 Q_3^6 + \frac{1}{18} Q_1^5 Q_2^3 Q_3^5 + \frac{1}{18} Q_1^5 Q_2^5 Q_3^3 - \frac{\sqrt{3}}{36} Q_1^5 Q_2^5 Q_3^8 - \frac{1}{24} Q_1^5 Q_2^6 Q_3^2 \nonumber \\
& & \mbox{} - \frac{1}{24} Q_1^5 Q_2^7 Q_3^1 - \frac{\sqrt{3}}{36} Q_1^5 Q_2^8 Q_3^5 + \frac{1}{24} Q_1^6 Q_2^1 Q_3^4 + \frac{1}{24} Q_1^6 Q_2^2 Q_3^5 + \frac{1}{18} Q_1^6 Q_2^3 Q_3^6 \nonumber \\
& & \mbox{} + \frac{1}{24} Q_1^6 Q_2^4 Q_3^1 + \frac{1}{24} Q_1^6 Q_2^5 Q_3^2 + \frac{1}{18} Q_1^6 Q_2^6 Q_3^3 + \frac{\sqrt{3}}{36} Q_1^6 Q_2^6 Q_3^8 + \frac{\sqrt{3}}{36} Q_1^6 Q_2^8 Q_3^6 \nonumber \\
& & \mbox{} + \frac{1}{24} Q_1^7 Q_2^1 Q_3^5 - \frac{1}{24} Q_1^7 Q_2^2 Q_3^4 + \frac{1}{18} Q_1^7 Q_2^3 Q_3^7 - \frac{1}{24} Q_1^7 Q_2^4 Q_3^2 + \frac{1}{24} Q_1^7 Q_2^5 Q_3^1 \nonumber \\
& & \mbox{} + \frac{1}{18} Q_1^7 Q_2^7 Q_3^3 + \frac{\sqrt{3}}{36} Q_1^7 Q_2^7 Q_3^8 + \frac{\sqrt{3}}{36} Q_1^7 Q_2^8 Q_3^7 - \frac16 Q_1^8 Q_2^3 Q_3^8 + \frac{\sqrt{3}}{18} Q_1^8 Q_2^4 Q_3^4 \nonumber \\
& & \mbox{} + \frac{\sqrt{3}}{18} Q_1^8 Q_2^5 Q_3^5 - \frac{\sqrt{3}}{18} Q_1^8 Q_2^6 Q_3^6 - \frac{\sqrt{3}}{18} Q_1^8 Q_2^7 Q_3^7 - \frac16 Q_1^8 Q_2^8 Q_3^3,
\end{eqnarray}

\begin{eqnarray}
[\tilde{\mathcal{P}}^{(64)} Q_1 Q_2 Q_3]^{388} & = & \frac{1}{126} Q_1^1 Q_2^1 Q_3^3 + \frac{1}{126} Q_1^1 Q_2^3 Q_3^1 + \frac{1}{126}
 Q_1^2 Q_2^2 Q_3^3 + \frac{1}{126} Q_1^2 Q_2^3 Q_3^2 + \frac{1}{126} Q_1^3 Q_2^1 Q_3^1 \nonumber \\
& & \mbox{} + \frac{1}{126} Q_1^3 Q_2^2 Q_3^2 + \frac{1}{42} Q_1^3 Q_2^3 Q_3^3 - \frac{5}{126} Q_1^3 Q_2^4 Q_3^4 - \frac{5}{126}
 Q_1^3 Q_2^5 Q_3^5 - \frac{5}{126} Q_1^3 Q_2^6 Q_3^6 \nonumber \\
& & \mbox{} - \frac{5}{126} Q_1^3 Q_2^7 Q_3^7 + \frac{5}{42} Q_1^3 Q_2^8 Q_3^8 - \frac{5}{126} Q_1^4 Q_2^3 Q_3^4 - \frac{5}{126} Q_1^4 Q_2^4 Q_3^3 - \frac{5\sqrt{3}}{126}5 Q_1^4 Q_2^4 Q_3^8 \nonumber \\
& & \mbox{} - \frac{5\sqrt{3}}{126}5 Q_1^4 Q_2^8 Q_3^4 - \frac{5}{126} Q_1^5 Q_2^3 Q_3^5 - \frac{5}{126} Q_1^5 Q_2^5 Q_3^3 - \frac{5\sqrt{3}}{126}5 Q_1^5 Q_2^5 Q_3^8 - \frac{5\sqrt{3}}{126}5 Q_1^5 Q_2^8 Q_3^5 \nonumber \\
& & \mbox{} - \frac{5}{126} Q_1^6 Q_2^3 Q_3^6 - \frac{5}{126} Q_1^6 Q_2^6 Q_3^3 + \frac{5\sqrt{3}}{126}5 Q_1^6 Q_2^6 Q_3^8 + \frac{5\sqrt{3}}{126}5 Q_1^6 Q_2^8 Q_3^6 - \frac{5}{126} Q_1^7 Q_2^3 Q_3^7 \nonumber \\
& & \mbox{} - \frac{5}{126} Q_1^7 Q_2^7 Q_3^3 + \frac{5\sqrt{3}}{126}5 Q_1^7 Q_2^7 Q_3^8 + \frac{5\sqrt{3}}{126}5 Q_1^7 Q_2^8 Q_3^7 + \frac{5}{42} Q_1^8 Q_2^3 Q_3^8 - \frac{5\sqrt{3}}{126}5 Q_1^8 Q_2^4 Q_3^4 \nonumber \\
& & \mbox{} - \frac{5\sqrt{3}}{126}5 Q_1^8 Q_2^5 Q_3^5 + \frac{5\sqrt{3}}{126}5 Q_1^8 Q_2^6 Q_3^6 + \frac{5\sqrt{3}}{126}5 Q_1^8 Q_2^7 Q_3^7 + \frac{5}{42} Q_1^8 Q_2^8 Q_3^3.
\end{eqnarray}

\subsection{$I = 2$}

\begin{eqnarray}
[\tilde{\mathcal{P}}^{(1)} Q_1 Q_2 Q_3]^{338} & = & \frac{1}{40} Q_1^1 Q_2^1 Q_3^8 + \frac{\sqrt{3}}{80} Q_1^1 Q_2^4 Q_3^6 + \frac{\sqrt{3}}{80} Q_1^1 Q_2^5 Q_3^7 + \frac{\sqrt{3}}{80} Q_1^1 Q_2^6 Q_3^4 + \frac{\sqrt{3}}{80} Q_1^1 Q_2^7 Q_3^5 \nonumber \\
& & \mbox{} + \frac{1}{40} Q_1^1 Q_2^8 Q_3^1 + \frac{1}{40} Q_1^2 Q_2^2 Q_3^8 - \frac{\sqrt{3}}{80} Q_1^2 Q_2^4 Q_3^7 + \frac{\sqrt{3}}{80} Q_1^2 Q_2^5 Q_3^6 + \frac{\sqrt{3}}{80} Q_1^2 Q_2^6 Q_3^5 \nonumber \\
& & \mbox{} - \frac{\sqrt{3}}{80} Q_1^2 Q_2^7 Q_3^4 + \frac{1}{40} Q_1^2 Q_2^8 Q_3^2 + \frac{1}{40} Q_1^3 Q_2^3 Q_3^8 + \frac{\sqrt{3}}{80} Q_1^3 Q_2^4 Q_3^4 + \frac{\sqrt{3}}{80} Q_1^3 Q_2^5 Q_3^5 \nonumber \\
& & \mbox{} - \frac{\sqrt{3}}{80} Q_1^3 Q_2^6 Q_3^6 - \frac{\sqrt{3}}{80} Q_1^3 Q_2^7 Q_3^7 + \frac{1}{40} Q_1^3 Q_2^8 Q_3^3 + \frac{\sqrt{3}}{80} Q_1^4 Q_2^1 Q_3^6 - \frac{\sqrt{3}}{80} Q_1^4 Q_2^2 Q_3^7 \nonumber \\
& & \mbox{} + \frac{\sqrt{3}}{80} Q_1^4 Q_2^3 Q_3^4 + \frac{\sqrt{3}}{80} Q_1^4 Q_2^4 Q_3^3 - \frac{1}{80} Q_1^4 Q_2^4 Q_3^8 + \frac{\sqrt{3}}{80} Q_1^4 Q_2^6 Q_3^1 - \frac{\sqrt{3}}{80} Q_1^4 Q_2^7 Q_3^2 \nonumber \\
& & \mbox{} - \frac{1}{80} Q_1^4 Q_2^8 Q_3^4 + \frac{\sqrt{3}}{80} Q_1^5 Q_2^1 Q_3^7 + \frac{\sqrt{3}}{80} Q_1^5 Q_2^2 Q_3^6 + \frac{\sqrt{3}}{80} Q_1^5 Q_2^3 Q_3^5 + \frac{\sqrt{3}}{80} Q_1^5 Q_2^5 Q_3^3 \nonumber \\
& & \mbox{} - \frac{1}{80} Q_1^5 Q_2^5 Q_3^8 + \frac{\sqrt{3}}{80} Q_1^5 Q_2^6 Q_3^2 + \frac{\sqrt{3}}{80} Q_1^5 Q_2^7 Q_3^1 - \frac{1}{80} Q_1^5 Q_2^8 Q_3^5 + \frac{\sqrt{3}}{80} Q_1^6 Q_2^1 Q_3^4 \nonumber \\
& & \mbox{} + \frac{\sqrt{3}}{80} Q_1^6 Q_2^2 Q_3^5 - \frac{\sqrt{3}}{80} Q_1^6 Q_2^3 Q_3^6 + \frac{\sqrt{3}}{80} Q_1^6 Q_2^4 Q_3^1 + \frac{\sqrt{3}}{80} Q_1^6 Q_2^5 Q_3^2 - \frac{\sqrt{3}}{80} Q_1^6 Q_2^6 Q_3^3 \nonumber \\
& & \mbox{} - \frac{1}{80} Q_1^6 Q_2^6 Q_3^8 - \frac{1}{80} Q_1^6 Q_2^8 Q_3^6 + \frac{\sqrt{3}}{80} Q_1^7 Q_2^1 Q_3^5 - \frac{\sqrt{3}}{80} Q_1^7 Q_2^2 Q_3^4 - \frac{\sqrt{3}}{80} Q_1^7 Q_2^3 Q_3^7 \nonumber \\
& & \mbox{} - \frac{\sqrt{3}}{80} Q_1^7 Q_2^4 Q_3^2 + \frac{\sqrt{3}}{80} Q_1^7 Q_2^5 Q_3^1 - \frac{\sqrt{3}}{80} Q_1^7 Q_2^7 Q_3^3 - \frac{1}{80} Q_1^7 Q_2^7 Q_3^8 - \frac{1}{80} Q_1^7 Q_2^8 Q_3^7 \nonumber \\
& & \mbox{} + \frac{1}{40} Q_1^8 Q_2^1 Q_3^1 + \frac{1}{40} Q_1^8 Q_2^2 Q_3^2 + \frac{1}{40} Q_1^8 Q_2^3 Q_3^3 - \frac{1}{80} Q_1^8 Q_2^4 Q_3^4 - \frac{1}{80} Q_1^8 Q_2^5 Q_3^5 \nonumber \\
& & \mbox{} - \frac{1}{80} Q_1^8 Q_2^6 Q_3^6 - \frac{1}{80} Q_1^8 Q_2^7 Q_3^7 - \frac{1}{40} Q_1^8 Q_2^8 Q_3^8,
\end{eqnarray}

\begin{eqnarray}
[\tilde{\mathcal{P}}^{(8)} Q_1 Q_2 Q_3]^{338} & = & \frac16 Q_1^1 Q_2^1 Q_3^8 + \frac{\sqrt{3}}{90} Q_1^1 Q_2^4 Q_3^6 + \frac{\sqrt{3}}{90} Q_1^1 Q_2^5 Q_3^7 + \frac{\sqrt{3}}{90} Q_1^1 Q_2^6 Q_3^4 + \frac{\sqrt{3}}{90} Q_1^1 Q_2^7 Q_3^5 \nonumber \\
& & \mbox{} - \frac{1}{30} Q_1^1 Q_2^8 Q_3^1 + \frac16 Q_1^2 Q_2^2 Q_3^8 - \frac{\sqrt{3}}{90} Q_1^2 Q_2^4 Q_3^7 + \frac{\sqrt{3}}{90} Q_1^2 Q_2^5 Q_3^6 + \frac{\sqrt{3}}{90} Q_1^2 Q_2^6 Q_3^5 \nonumber \\
& & \mbox{} - \frac{\sqrt{3}}{90} Q_1^2 Q_2^7 Q_3^4 - \frac{1}{30} Q_1^2 Q_2^8 Q_3^2 + \frac16 Q_1^3 Q_2^3 Q_3^8 + \frac{\sqrt{3}}{90} Q_1^3 Q_2^4 Q_3^4 + \frac{\sqrt{3}}{90} Q_1^3 Q_2^5 Q_3^5 \nonumber \\
& & \mbox{} - \frac{\sqrt{3}}{90} Q_1^3 Q_2^6 Q_3^6 - \frac{\sqrt{3}}{90} Q_1^3 Q_2^7 Q_3^7 - \frac{1}{30} Q_1^3 Q_2^8 Q_3^3 + \frac{\sqrt{3}}{90} Q_1^4 Q_2^1 Q_3^6 - \frac{\sqrt{3}}{90} Q_1^4 Q_2^2 Q_3^7 \nonumber \\
& & \mbox{} + \frac{\sqrt{3}}{90} Q_1^4 Q_2^3 Q_3^4 - \frac{\sqrt{3}}{45} Q_1^4 Q_2^4 Q_3^3 + \frac{1}{10} Q_1^4 Q_2^4 Q_3^8 - \frac{\sqrt{3}}{45} Q_1^4 Q_2^6 Q_3^1 + \frac{\sqrt{3}}{45} Q_1^4 Q_2^7 Q_3^2 \nonumber \\
& & \mbox{} + \frac{\sqrt{3}}{90} Q_1^5 Q_2^1 Q_3^7 + \frac{\sqrt{3}}{90} Q_1^5 Q_2^2 Q_3^6 + \frac{\sqrt{3}}{90} Q_1^5 Q_2^3 Q_3^5 - \frac{\sqrt{3}}{45} Q_1^5 Q_2^5 Q_3^3 + \frac{1}{10} Q_1^5 Q_2^5 Q_3^8 \nonumber \\
& & \mbox{} - \frac{\sqrt{3}}{45} Q_1^5 Q_2^6 Q_3^2 - \frac{\sqrt{3}}{45} Q_1^5 Q_2^7 Q_3^1 + \frac{\sqrt{3}}{90} Q_1^6 Q_2^1 Q_3^4 + \frac{\sqrt{3}}{90} Q_1^6 Q_2^2 Q_3^5 - \frac{\sqrt{3}}{90} Q_1^6 Q_2^3 Q_3^6 \nonumber \\
& & \mbox{} - \frac{\sqrt{3}}{45} Q_1^6 Q_2^4 Q_3^1 - \frac{\sqrt{3}}{45} Q_1^6 Q_2^5 Q_3^2 + \frac{\sqrt{3}}{45} Q_1^6 Q_2^6 Q_3^3 + \frac{1}{10} Q_1^6 Q_2^6 Q_3^8 + \frac{\sqrt{3}}{90} Q_1^7 Q_2^1 Q_3^5 \nonumber \\
& & \mbox{} - \frac{\sqrt{3}}{90} Q_1^7 Q_2^2 Q_3^4 - \frac{\sqrt{3}}{90} Q_1^7 Q_2^3 Q_3^7 + \frac{\sqrt{3}}{45} Q_1^7 Q_2^4 Q_3^2 - \frac{\sqrt{3}}{45} Q_1^7 Q_2^5 Q_3^1 + \frac{\sqrt{3}}{45} Q_1^7 Q_2^7 Q_3^3 \nonumber \\
& & \mbox{} + \frac{1}{10} Q_1^7 Q_2^7 Q_3^8 - \frac{1}{30} Q_1^8 Q_2^1 Q_3^1 - \frac{1}{30} Q_1^8 Q_2^2 Q_3^2 - \frac{1}{30} Q_1^8 Q_2^3 Q_3^3 + \frac{1}{10} Q_1^8 Q_2^8 Q_3^8,
\end{eqnarray}

\begin{equation}
[\tilde{\mathcal{P}}^{(10+\overline{10})} Q_1 Q_2 Q_3]^{338} = 0,
\end{equation}

\begin{eqnarray}
[\tilde{\mathcal{P}}^{(27)} Q_1 Q_2 Q_3]^{338} & = & \frac{1}{56} Q_1^1 Q_2^1 Q_3^8 - \frac{29\sqrt{3}}{560} Q_1^1 Q_2^4 Q_3^6 - \frac{29\sqrt{3}}{560} Q_1^1 Q_2^5 Q_3^7 - \frac{29\sqrt{3}}{560} Q_1^1 Q_2^6 Q_3^4 - \frac{29\sqrt{3}}{560} Q_1^1 Q_2^7 Q_3^5 \nonumber \\
& & \mbox{} - \frac{9}{280} Q_1^1 Q_2^8 Q_3^1 + \frac{1}{56} Q_1^2 Q_2^2 Q_3^8 + \frac{29\sqrt{3}}{560} Q_1^2 Q_2^4 Q_3^7 - \frac{29\sqrt{3}}{560} Q_1^2 Q_2^5 Q_3^6 - \frac{29\sqrt{3}}{560} Q_1^2 Q_2^6 Q_3^5 \nonumber \\
& & \mbox{} + \frac{29\sqrt{3}}{560} Q_1^2 Q_2^7 Q_3^4 - \frac{9}{280} Q_1^2 Q_2^8 Q_3^2 + \frac38 Q_1^3 Q_2^3 Q_3^8 + \frac{3\sqrt{3}}{80} Q_1^3 Q_2^4 Q_3^4 + \frac{3\sqrt{3}}{80} Q_1^3 Q_2^5 Q_3^5 \nonumber \\
& & \mbox{} - \frac{3\sqrt{3}}{80} Q_1^3 Q_2^6 Q_3^6 - \frac{3\sqrt{3}}{80} Q_1^3 Q_2^7 Q_3^7 + \frac{3}{40} Q_1^3 Q_2^8 Q_3^3 - \frac{29\sqrt{3}}{560} Q_1^4 Q_2^1 Q_3^6 + \frac{29\sqrt{3}}{560} Q_1^4 Q_2^2 Q_3^7 \nonumber \\
& & \mbox{} + \frac{3\sqrt{3}}{80} Q_1^4 Q_2^3 Q_3^4 - \frac{\sqrt{3}}{80} Q_1^4 Q_2^4 Q_3^3 - \frac{41}{560} Q_1^4 Q_2^4 Q_3^8 + \frac{13\sqrt{3}}{560} Q_1^4 Q_2^6 Q_3^1 - \frac{13\sqrt{3}}{560} Q_1^4 Q_2^7 Q_3^2 \nonumber \\
& & \mbox{} + \frac{3}{112} Q_1^4 Q_2^8 Q_3^4 - \frac{29\sqrt{3}}{560} Q_1^5 Q_2^1 Q_3^7 - \frac{29\sqrt{3}}{560} Q_1^5 Q_2^2 Q_3^6 + \frac{3\sqrt{3}}{80} Q_1^5 Q_2^3 Q_3^5 - \frac{\sqrt{3}}{80} Q_1^5 Q_2^5 Q_3^3 \nonumber \\
& & \mbox{} - \frac{41}{560} Q_1^5 Q_2^5 Q_3^8 + \frac{13\sqrt{3}}{560} Q_1^5 Q_2^6 Q_3^2 + \frac{13\sqrt{3}}{560} Q_1^5 Q_2^7 Q_3^1 + \frac{3}{112} Q_1^5 Q_2^8 Q_3^5 - \frac{29\sqrt{3}}{560} Q_1^6 Q_2^1 Q_3^4 \nonumber \\
& & \mbox{} - \frac{29\sqrt{3}}{560} Q_1^6 Q_2^2 Q_3^5 - \frac{3\sqrt{3}}{80} Q_1^6 Q_2^3 Q_3^6 + \frac{13\sqrt{3}}{560} Q_1^6 Q_2^4 Q_3^1 + \frac{13\sqrt{3}}{560} Q_1^6 Q_2^5 Q_3^2 + \frac{\sqrt{3}}{80} Q_1^6 Q_2^6 Q_3^3 \nonumber \\
& & \mbox{} - \frac{41}{560} Q_1^6 Q_2^6 Q_3^8 + \frac{3}{112} Q_1^6 Q_2^8 Q_3^6 - \frac{29\sqrt{3}}{560} Q_1^7 Q_2^1 Q_3^5 +\frac{29\sqrt{3}}{560} Q_1^7 Q_2^2 Q_3^4 - \frac{3\sqrt{3}}{80} Q_1^7 Q_2^3 Q_3^7 \nonumber \\
& & \mbox{} - \frac{13\sqrt{3}}{560} Q_1^7 Q_2^4 Q_3^2 + \frac{13\sqrt{3}}{560} Q_1^7 Q_2^5 Q_3^1 + \frac{\sqrt{3}}{80} Q_1^7 Q_2^7 Q_3^3 - \frac{41}{560} Q_1^7 Q_2^7 Q_3^8 + \frac{3}{112} Q_1^7 Q_2^8 Q_3^7 \nonumber \\
& & \mbox{} - \frac{9}{280} Q_1^8 Q_2^1 Q_3^1 - \frac{9}{280} Q_1^8 Q_2^2 Q_3^2 + \frac{3}{40} Q_1^8 Q_2^3 Q_3^3 + \frac{3}{112} Q_1^8 Q_2^4 Q_3^4 + \frac{3}{112} Q_1^8 Q_2^5 Q_3^5 \nonumber \\
& & \mbox{} + \frac{3}{112} Q_1^8 Q_2^6 Q_3^6 + \frac{3}{112} Q_1^8 Q_2^7 Q_3^7 - \frac{33}{280} Q_1^8 Q_2^8 Q_3^8,
\end{eqnarray}

\begin{eqnarray}
[\tilde{\mathcal{P}}^{(35+\overline{35})} Q_1 Q_2 Q_3]^{338} & = & - \frac16 Q_1^1 Q_2^1 Q_3^8 + \frac{\sqrt{3}}{72} Q_1^1 Q_2^4 Q_3^6 + \frac{\sqrt{3}}{72} Q_1^1 Q_2^5 Q_3^7 + \frac{\sqrt{3}}{72} Q_1^1 Q_2^6 Q_3^4 + \frac{\sqrt{3}}{72} Q_1^1 Q_2^7 Q_3^5 \nonumber \\
& & \mbox{} + \frac{1}{12} Q_1^1 Q_2^8 Q_3^1 - \frac16 Q_1^2 Q_2^2 Q_3^8 - \frac{\sqrt{3}}{72} Q_1^2 Q_2^4 Q_3^7 + \frac{\sqrt{3}}{72} Q_1^2 Q_2^5 Q_3^6 + \frac{\sqrt{3}}{72} Q_1^2 Q_2^6 Q_3^5 \nonumber \\
& & \mbox{} - \frac{\sqrt{3}}{72} Q_1^2 Q_2^7 Q_3^4 + \frac{1}{12} Q_1^2 Q_2^8 Q_3^2 + \frac13 Q_1^3 Q_2^3 Q_3^8 - \frac{\sqrt{3}}{36} Q_1^3 Q_2^4 Q_3^4 - \frac{\sqrt{3}}{36} Q_1^3 Q_2^5 Q_3^5 \nonumber \\
& & \mbox{} + \frac{\sqrt{3}}{36} Q_1^3 Q_2^6 Q_3^6 + \frac{\sqrt{3}}{36} Q_1^3 Q_2^7 Q_3^7 - \frac16 Q_1^3 Q_2^8 Q_3^3 + \frac{\sqrt{3}}{72} Q_1^4 Q_2^1 Q_3^6 - \frac{\sqrt{3}}{72} Q_1^4 Q_2^2 Q_3^7 \nonumber \\
& & \mbox{} - \frac{\sqrt{3}}{36} Q_1^4 Q_2^3 Q_3^4 + \frac{\sqrt{3}}{18} Q_1^4 Q_2^4 Q_3^3 - \frac{\sqrt{3}}{36} Q_1^4 Q_2^6 Q_3^1 + \frac{\sqrt{3}}{36} Q_1^4 Q_2^7 Q_3^2 + \frac{\sqrt{3}}{72} Q_1^5 Q_2^1 Q_3^7 \nonumber \\
& & \mbox{} + \frac{\sqrt{3}}{72} Q_1^5 Q_2^2 Q_3^6 - \frac{\sqrt{3}}{36} Q_1^5 Q_2^3 Q_3^5 + \frac{\sqrt{3}}{18} Q_1^5 Q_2^5 Q_3^3 - \frac{\sqrt{3}}{36} Q_1^5 Q_2^6 Q_3^2 - \frac{\sqrt{3}}{36} Q_1^5 Q_2^7 Q_3^1 \nonumber \\
& & \mbox{} + \frac{\sqrt{3}}{72} Q_1^6 Q_2^1 Q_3^4 + \frac{\sqrt{3}}{72} Q_1^6 Q_2^2 Q_3^5 + \frac{\sqrt{3}}{36} Q_1^6 Q_2^3 Q_3^6 - \frac{\sqrt{3}}{36} Q_1^6 Q_2^4 Q_3^1 - \frac{\sqrt{3}}{36} Q_1^6 Q_2^5 Q_3^2 \nonumber \\
& & \mbox{} - \frac{\sqrt{3}}{18} Q_1^6 Q_2^6 Q_3^3 + \frac{\sqrt{3}}{72} Q_1^7 Q_2^1 Q_3^5 - \frac{\sqrt{3}}{72} Q_1^7 Q_2^2 Q_3^4 + \frac{\sqrt{3}}{36} Q_1^7 Q_2^3 Q_3^7 + \frac{\sqrt{3}}{36} Q_1^7 Q_2^4 Q_3^2 \nonumber \\
& & \mbox{} - \frac{\sqrt{3}}{36} Q_1^7 Q_2^5 Q_3^1 - \frac{\sqrt{3}}{18} Q_1^7 Q_2^7 Q_3^3 + \frac{1}{12} Q_1^8 Q_2^1 Q_3^1 + \frac{1}{12} Q_1^8 Q_2^2 Q_3^2 - \frac16 Q_1^8 Q_2^3 Q_3^3,
\end{eqnarray}

\begin{eqnarray}
[\tilde{\mathcal{P}}^{(64)} Q_1 Q_2 Q_3]^{338} & = & - \frac{3}{70} Q_1^1 Q_2^1 Q_3^8 + \frac{\sqrt{3}}{70} Q_1^1 Q_2^4 Q_3^6 + \frac{\sqrt{3}}{70} Q_1^1 Q_2^5 Q_3^7 + \frac{\sqrt{3}}{70} Q_1^1 Q_2^6 Q_3^4 + \frac{\sqrt{3}}{70} Q_1^1 Q_2^7 Q_3^5 \nonumber \\
& & \mbox{} - \frac{3}{70} Q_1^1 Q_2^8 Q_3^1 - \frac{3}{70} Q_1^2 Q_2^2 Q_3^8 - \frac{\sqrt{3}}{70} Q_1^2 Q_2^4 Q_3^7 + \frac{\sqrt{3}}{70} Q_1^2 Q_2^5 Q_3^6 + \frac{\sqrt{3}}{70} Q_1^2 Q_2^6 Q_3^5 \nonumber \\
& & \mbox{} - \frac{\sqrt{3}}{70} Q_1^2 Q_2^7 Q_3^4 - \frac{3}{70} Q_1^2 Q_2^8 Q_3^2 + \frac{1}{10} Q_1^3 Q_2^3 Q_3^8 - \frac{\sqrt{3}}{30} Q_1^3 Q_2^4 Q_3^4 - \frac{\sqrt{3}}{30} Q_1^3 Q_2^5 Q_3^5 \nonumber \\
& & \mbox{} + \frac{\sqrt{3}}{30} Q_1^3 Q_2^6 Q_3^6 + \frac{\sqrt{3}}{30} Q_1^3 Q_2^7 Q_3^7 + \frac{1}{10} Q_1^3 Q_2^8 Q_3^3 +\frac{\sqrt{3}}{70} Q_1^4 Q_2^1 Q_3^6 - \frac{\sqrt{3}}{70} Q_1^4 Q_2^2 Q_3^7 \nonumber \\
& & \mbox{} - \frac{\sqrt{3}}{30} Q_1^4 Q_2^3 Q_3^4 - \frac{\sqrt{3}}{30} Q_1^4 Q_2^4 Q_3^3 - \frac{1}{70} Q_1^4 Q_2^4 Q_3^8 + \frac{\sqrt{3}}{70} Q_1^4 Q_2^6 Q_3^1 - \frac{\sqrt{3}}{70} Q_1^4 Q_2^7 Q_3^2 \nonumber \\
& & \mbox{} - \frac{1}{70} Q_1^4 Q_2^8 Q_3^4 + \frac{\sqrt{3}}{70} Q_1^5 Q_2^1 Q_3^7 + \frac{\sqrt{3}}{70} Q_1^5 Q_2^2 Q_3^6 - \frac{\sqrt{3}}{30} Q_1^5 Q_2^3 Q_3^5 - \frac{\sqrt{3}}{30} Q_1^5 Q_2^5 Q_3^3 \nonumber \\
& & \mbox{} - \frac{1}{70} Q_1^5 Q_2^5 Q_3^8 + \frac{\sqrt{3}}{70} Q_1^5 Q_2^6 Q_3^2 + \frac{\sqrt{3}}{70} Q_1^5 Q_2^7 Q_3^1 - \frac{1}{70} Q_1^5 Q_2^8 Q_3^5 + \frac{\sqrt{3}}{70} Q_1^6 Q_2^1 Q_3^4 \nonumber \\
& & \mbox{} + \frac{\sqrt{3}}{70} Q_1^6 Q_2^2 Q_3^5 + \frac{\sqrt{3}}{30} Q_1^6 Q_2^3 Q_3^6 + \frac{\sqrt{3}}{70} Q_1^6 Q_2^4 Q_3^1 + \frac{\sqrt{3}}{70} Q_1^6 Q_2^5 Q_3^2 + \frac{\sqrt{3}}{30} Q_1^6 Q_2^6 Q_3^3 \nonumber \\
& & \mbox{} - \frac{1}{70} Q_1^6 Q_2^6 Q_3^8 - \frac{1}{70} Q_1^6 Q_2^8 Q_3^6 + \frac{\sqrt{3}}{70} Q_1^7 Q_2^1 Q_3^5 - \frac{\sqrt{3}}{70} Q_1^7 Q_2^2 Q_3^4 + \frac{\sqrt{3}}{30} Q_1^7 Q_2^3 Q_3^7 \nonumber \\
& & \mbox{} - \frac{\sqrt{3}}{70} Q_1^7 Q_2^4 Q_3^2 + \frac{\sqrt{3}}{70} Q_1^7 Q_2^5 Q_3^1 + \frac{\sqrt{3}}{30} Q_1^7 Q_2^7 Q_3^3 - \frac{1}{70} Q_1^7 Q_2^7 Q_3^8 - \frac{1}{70} Q_1^7 Q_2^8 Q_3^7 \nonumber \\
& & \mbox{} - \frac{3}{70} Q_1^8 Q_2^1 Q_3^1 - \frac{3}{70} Q_1^8 Q_2^2 Q_3^2 + \frac{1}{10} Q_1^8 Q_2^3 Q_3^3 - \frac{1}{70} Q_1^8 Q_2^4 Q_3^4 - \frac{1}{70} Q_1^8 Q_2^5 Q_3^5 \nonumber \\
& & \mbox{} - \frac{1}{70} Q_1^8 Q_2^6 Q_3^6 - \frac{1}{70} Q_1^8 Q_2^7 Q_3^7 + \frac{3}{70} Q_1^8 Q_2^8 Q_3^8.
\end{eqnarray}

\subsection{$I = 3$}

\begin{equation}
[\tilde{\mathcal{P}}^{(1)} Q_1 Q_2 Q_3]^{333} = 0,
\end{equation}

\begin{eqnarray}
[\tilde{\mathcal{P}}^{(8)} Q_1 Q_2 Q_3]^{333} & = & \frac{1}{10} Q_1^1 Q_2^1 Q_3^3 + \frac{1}{10} Q_1^1 Q_2^3 Q_3^1 + \frac{1}{10} Q_1^2 Q_2^2 Q_3^3 + \frac{1}{10} Q_1^2 Q_2^3 Q_3^2 + \frac{1}{10} Q_1^3 Q_2^1 Q_3^1 \nonumber \\
& & \mbox{} + \frac{1}{10} Q_1^3 Q_2^2 Q_3^2 + \frac{3}{10} Q_1^3 Q_2^3 Q_3^3 + \frac{1}{10} Q_1^3 Q_2^4 Q_3^4 + \frac{1}{10} Q_1^3 Q_2^5 Q_3^5 + \frac{1}{10} Q_1^3 Q_2^6 Q_3^6 \nonumber \\
& & \mbox{} + \frac{1}{10} Q_1^3 Q_2^7 Q_3^7 + \frac{1}{10} Q_1^3 Q_2^8 Q_3^8 + \frac{1}{10} Q_1^4 Q_2^3 Q_3^4 + \frac{1}{10} Q_1^4 Q_2^4 Q_3^3 + \frac{1}{10} Q_1^5 Q_2^3 Q_3^5 \nonumber \\
& & \mbox{} + \frac{1}{10} Q_1^5 Q_2^5 Q_3^3 + \frac{1}{10} Q_1^6 Q_2^3 Q_3^6 + \frac{1}{10} Q_1^6 Q_2^6 Q_3^3 + \frac{1}{10} Q_1^7 Q_2^3 Q_3^7 + \frac{1}{10} Q_1^7 Q_2^7 Q_3^3 \nonumber \\
& & \mbox{} + \frac{1}{10} Q_1^8 Q_2^3 Q_3^8 + \frac{1}{10} Q_1^8 Q_2^8 Q_3^3,
\end{eqnarray}

\begin{eqnarray}
[\tilde{\mathcal{P}}^{(10+\overline{10})} Q_1 Q_2 Q_3]^{333} & = & \frac{1}{18} Q_1^1 Q_2^1 Q_3^3 + \frac{1}{18} Q_1^1 Q_2^3 Q_3^1 + \frac{1}{18}
 Q_1^2 Q_2^2 Q_3^3 + \frac{1}{18} Q_1^2 Q_2^3 Q_3^2 + \frac{1}{18} Q_1^3 Q_2^1 Q_3^1 \nonumber \\
& & \mbox{} + \frac{1}{18} Q_1^3 Q_2^2 Q_3^2 + \frac16 Q_1^3 Q_2^3 Q_3^3 - \frac{1}{36} Q_1^3 Q_2^4 Q_3^4 - \frac{1}{36} Q_1^3 Q_2^5 Q_3^5 - \frac{1}{36} Q_1^3 Q_2^6 Q_3^6 \nonumber \\
& & \mbox{} - \frac{1}{36} Q_1^3 Q_2^7 Q_3^7 - \frac16 Q_1^3 Q_2^8 Q_3^8 - \frac{1}{36} Q_1^4 Q_2^3 Q_3^4 - \frac{1}{36} Q_1^4 Q_2^4 Q_3^3 - \frac{\sqrt{3}}{36} Q_1^4 Q_2^4 Q_3^8 \nonumber \\
& & \mbox{} - \frac{\sqrt{3}}{36} Q_1^4 Q_2^8 Q_3^4 - \frac{1}{36} Q_1^5 Q_2^3 Q_3^5 - \frac{1}{36} Q_1^5 Q_2^5 Q_3^3 - \frac{\sqrt{3}}{36} Q_1^5 Q_2^5 Q_3^8 - \frac{\sqrt{3}}{36} Q_1^5 Q_2^8 Q_3^5 \nonumber \\
& & \mbox{} - \frac{1}{36} Q_1^6 Q_2^3 Q_3^6 - \frac{1}{36} Q_1^6 Q_2^6 Q_3^3 + \frac{\sqrt{3}}{36} Q_1^6 Q_2^6 Q_3^8 + \frac{\sqrt{3}}{36} Q_1^6 Q_2^8 Q_3^6 - \frac{1}{36} Q_1^7 Q_2^3 Q_3^7 \nonumber \\
& & \mbox{} - \frac{1}{36} Q_1^7 Q_2^7 Q_3^3 + \frac{\sqrt{3}}{36} Q_1^7 Q_2^7 Q_3^8 + \frac{\sqrt{3}}{36} Q_1^7 Q_2^8 Q_3^7 - \frac16 Q_1^8 Q_2^3 Q_3^8 - \frac{\sqrt{3}}{36} Q_1^8 Q_2^4 Q_3^4 \nonumber \\
& & \mbox{} - \frac{\sqrt{3}}{36} Q_1^8 Q_2^5 Q_3^5 + \frac{\sqrt{3}}{36} Q_1^8 Q_2^6 Q_3^6 + \frac{\sqrt{3}}{36} Q_1^8 Q_2^7 Q_3^7 - \frac16 Q_1^8 Q_2^8 Q_3^3,
\end{eqnarray}

\begin{eqnarray}
[\tilde{\mathcal{P}}^{(27)} Q_1 Q_2 Q_3]^{333} & = & \frac{3}{70} Q_1^1 Q_2^1 Q_3^3 + \frac{3}{70} Q_1^1 Q_2^3 Q_3^1 + \frac{3}{70} Q_1^2 Q_2^2 Q_3^3 + \frac{3}{70} Q_1^2 Q_2^3 Q_3^2 + \frac{3}{70} Q_1^3 Q_2^1 Q_3^1 \nonumber \\
& & \mbox{} + \frac{3}{70} Q_1^3 Q_2^2 Q_3^2 + \frac{9}{70} Q_1^3 Q_2^3 Q_3^3 - \frac{9}{140} Q_1^3 Q_2^4 Q_3^4 - \frac{9}{140} Q_1^3 Q_2^5 Q_3^5 - \frac{9}{140} Q_1^3 Q_2^6 Q_3^6 \nonumber \\
& & \mbox{} - \frac{9}{140} Q_1^3 Q_2^7 Q_3^7 + \frac{3}{70} Q_1^3 Q_2^8 Q_3^8 - \frac{9}{140} Q_1^4 Q_2^3 Q_3^4 - \frac{9}{140} Q_1^4 Q_2^4 Q_3^3 + \frac{\sqrt{3}}{28} Q_1^4 Q_2^4 Q_3^8 \nonumber \\
& & \mbox{} + \frac{\sqrt{3}}{28} Q_1^4 Q_2^8 Q_3^4 - \frac{9}{140} Q_1^5 Q_2^3 Q_3^5 - \frac{9}{140} Q_1^5 Q_2^5 Q_3^3 + \frac{\sqrt{3}}{28} Q_1^5 Q_2^5 Q_3^8 + \frac{\sqrt{3}}{28} Q_1^5 Q_2^8 Q_3^5 \nonumber \\
& & \mbox{} - \frac{9}{140} Q_1^6 Q_2^3 Q_3^6 - \frac{9}{140} Q_1^6 Q_2^6 Q_3^3 - \frac{\sqrt{3}}{28} Q_1^6 Q_2^6 Q_3^8 - \frac{\sqrt{3}}{28} Q_1^6 Q_2^8 Q_3^6 - \frac{9}{140} Q_1^7 Q_2^3 Q_3^7 \nonumber \\
& & \mbox{} - \frac{9}{140} Q_1^7 Q_2^7 Q_3^3 - \frac{\sqrt{3}}{28} Q_1^7 Q_2^7 Q_3^8 - \frac{\sqrt{3}}{28} Q_1^7 Q_2^8 Q_3^7 + \frac{3}{70} Q_1^8 Q_2^3 Q_3^8 + \frac{\sqrt{3}}{28} Q_1^8 Q_2^4 Q_3^4 \nonumber \\
& & \mbox{} + \frac{\sqrt{3}}{28} Q_1^8 Q_2^5 Q_3^5 - \frac{\sqrt{3}}{28} Q_1^8 Q_2^6 Q_3^6 - \frac{\sqrt{3}}{28} Q_1^8 Q_2^7 Q_3^7 + \frac{3}{70} Q_1^8 Q_2^8 Q_3^3,
\end{eqnarray}

\begin{equation}
[\tilde{\mathcal{P}}^{(35+\overline{35})} Q_1 Q_2 Q_3]^{333} = 0,
\end{equation}

\begin{eqnarray}
[\tilde{\mathcal{P}}^{(64)} Q_1 Q_2 Q_3]^{333} & = & - \frac{25}{126} Q_1^1 Q_2^1 Q_3^3 - \frac{25}{126} Q_1^1 Q_2^3 Q_3^1 - \frac{25}{126}
 Q_1^2 Q_2^2 Q_3^3 - \frac{25}{126} Q_1^2 Q_2^3 Q_3^2 - \frac{25}{126} Q_1^3 Q_2^1 Q_3^1 \nonumber \\
& & \mbox{} - \frac{25}{126} Q_1^3 Q_2^2 Q_3^2 + \frac{17}{42} Q_1^3 Q_2^3 Q_3^3 - \frac{1}{126} Q_1^3 Q_2^4 Q_3^4 - \frac{1}{126}
 Q_1^3 Q_2^5 Q_3^5 - \frac{1}{126} Q_1^3 Q_2^6 Q_3^6 \nonumber \\
& & \mbox{} - \frac{1}{126} Q_1^3 Q_2^7 Q_3^7 + \frac{1}{42} Q_1^3 Q_2^8 Q_3^8 - \frac{1}{126} Q_1^4 Q_2^3 Q_3^4 - \frac{1}{126} Q_1^4 Q_2^4 Q_3^3 - \frac{\sqrt{3}}{126} Q_1^4 Q_2^4 Q_3^8 \nonumber \\
& & \mbox{} - \frac{\sqrt{3}}{126} Q_1^4 Q_2^8 Q_3^4 - \frac{1}{126} Q_1^5 Q_2^3 Q_3^5 - \frac{1}{126} Q_1^5 Q_2^5 Q_3^3 - \frac{\sqrt{3}}{126} Q_1^5 Q_2^5 Q_3^8 - \frac{\sqrt{3}}{126} Q_1^5 Q_2^8 Q_3^5 \nonumber \\
& & \mbox{} - \frac{1}{126} Q_1^6 Q_2^3 Q_3^6 - \frac{1}{126} Q_1^6 Q_2^6 Q_3^3 + \frac{\sqrt{3}}{126} Q_1^6 Q_2^6 Q_3^8 + \frac{\sqrt{3}}{126} Q_1^6 Q_2^8 Q_3^6 - \frac{1}{126} Q_1^7 Q_2^3 Q_3^7 \nonumber \\
& & \mbox{} - \frac{1}{126} Q_1^7 Q_2^7 Q_3^3 + \frac{\sqrt{3}}{126} Q_1^7 Q_2^7 Q_3^8 + \frac{\sqrt{3}}{126} Q_1^7 Q_2^8 Q_3^7 + \frac{1}{42} Q_1^8 Q_2^3 Q_3^8 - \frac{\sqrt{3}}{126} Q_1^8 Q_2^4 Q_3^4 \nonumber \\
& & \mbox{} - \frac{\sqrt{3}}{126} Q_1^8 Q_2^5 Q_3^5 + \frac{\sqrt{3}}{126} Q_1^8 Q_2^6 Q_3^6 + \frac{\sqrt{3}}{126} Q_1^8 Q_2^7 Q_3^7 + \frac{1}{42} Q_1^8 Q_2^8 Q_3^3.
\end{eqnarray}

It is straightforward to prove that
\begin{eqnarray}
& & [\tilde{\mathcal{P}}^{(1)} Q_1 Q_2 Q_3]^{a_1a_2a_3} + [\tilde{\mathcal{P}}^{(8)} Q_1 Q_2 Q_3]^{a_1a_2a_3} + [\tilde{\mathcal{P}}^{(10+\overline{10})} Q_1 Q_2 Q_3]^{a_1a_2a_3} + [\tilde{\mathcal{P}}^{(27)} Q_1 Q_2 Q_3]^{a_1a_2a_3} \nonumber \\
& & \mbox{\hglue1.0truecm} + [\tilde{\mathcal{P}}^{(35+\overline{35})} Q_1 Q_2 Q_3]^{a_1a_2a_3} + [\tilde{\mathcal{P}}^{(64)} Q_1 Q_2 Q_3]^{a_1a_2a_3} = Q_1^{a_1} Q_2^{a_2} Q_3^{a_3},
\end{eqnarray}
according to the properties satisfied by projection operators.

Structures like $[\tilde{\mathcal{P}}^{(m)} \{Q_1,\{Q_2,Q_3\}\}]^{a_1a_2a_3}$ can easily be obtained from the expressions listed above.

\section{\label{app:fullmass}Full expressions for baryon masses}

The full theoretical expressions for the baryon masses can be expressed in terms of the 21 free operator coefficients required in the analysis. The expressions read
\begin{eqnarray}
M_{n} & = & N_c \tilde{m}_1^{1,0} + \frac{3}{4N_c} \tilde{m}_2^{1,0} - \frac12 \tilde{m}_1^{8,1} - \frac{5}{4N_c} \tilde{m}_2^{8,1} - \frac{3}{4N_c^2} \tilde{m}_3^{8,1} + \frac12 \tilde{m}_1^{8,0} + \frac{1}{4N_c} \tilde{m}_2^{8,0} + \frac{3}{4N_c^2} \tilde{m}_3^{8,0} \nonumber \\
& & \mbox{} + \frac{1}{2N_c^2} \tilde{m}_1^{10 + \overline{10},1} + \frac{1}{20N_c} \tilde{m}_1^{27,2} + \frac{1}{20N_c^2} \tilde{m}_2^{27,2} - \frac{1}{5N_c} \tilde{m}_1^{27,1} - \frac{1}{5N_c^2} \tilde{m}_2^{27,1} + \frac{9}{20N_c} \tilde{m}_1^{27,0} \nonumber \\
& & \mbox{} + \frac{9}{20N_c^2} \tilde{m}_2^{27,0},
\end{eqnarray}
\begin{eqnarray}
M_{p} & = & N_c \tilde{m}_1^{1,0} + \frac{3}{4N_c} \tilde{m}_2^{1,0} + \frac12 \tilde{m}_1^{8,1} + \frac{5}{4N_c} \tilde{m}_2^{8,1} + \frac{3}{4N_c^2} \tilde{m}_3^{8,1} + \frac12 \tilde{m}_1^{8,0} + \frac{1}{4N_c} \tilde{m}_2^{8,0} + \frac{3}{4N_c^2} \tilde{m}_3^{8,0} \nonumber \\
& & \mbox{} - \frac{1}{2N_c^2} \tilde{m}_1^{10 + \overline{10},1} + \frac{1}{20N_c} \tilde{m}_1^{27,2} + \frac{1}{20N_c^2} \tilde{m}_2^{27,2} + \frac{1}{5N_c} \tilde{m}_1^{27,1} + \frac{1}{5N_c^2} \tilde{m}_2^{27,1} + \frac{9}{20N_c} \tilde{m}_1^{27,0} \nonumber \\
& & \mbox{} + \frac{9}{20N_c^2} \tilde{m}_2^{27,0},
\end{eqnarray}
\begin{eqnarray}
M_{\Sigma^+} & = & N_c \tilde{m}_1^{1,0} + \frac{3}{4N_c} \tilde{m}_2^{1,0} + \tilde{m}_1^{8,1} + \frac{1}{N_c} \tilde{m}_2^{8,1} + \frac{3}{2N_c^2} \tilde{m}_3^{8,1} + \frac{1}{2N_c} \tilde{m}_2^{8,0} + \frac{1}{2N_c^2} \tilde{m}_1^{10 + \overline{10},1} \nonumber \\
& & \mbox{} + \frac{13}{20N_c} \tilde{m}_1^{27,2} + \frac{13}{20N_c^2} \tilde{m}_2^{27,2} - \frac{3}{20N_c} \tilde{m}_1^{27,0} - \frac{3}{20N_c^2} \tilde{m}_2^{27,0},
\end{eqnarray}
\begin{equation}
M_{\Sigma^0} = N_c \tilde{m}_1^{1,0} + \frac{3}{4N_c} \tilde{m}_2^{1,0} + \frac{1}{2N_c} \tilde{m}_2^{8,0} - \frac{27}{20N_c} \tilde{m}_1^{27,2} - \frac{27}{20N_c^2} \tilde{m}_2^{27,2} - \frac{3}{20N_c} \tilde{m}_1^{27,0} - \frac{3}{20N_c^2} \tilde{m}_2^{27,0},
\end{equation}
\begin{eqnarray}
M_{\Sigma^-} & = & N_c \tilde{m}_1^{1,0} + \frac{3}{4N_c} \tilde{m}_2^{1,0} - \tilde{m}_1^{8,1} - \frac{1}{N_c} \tilde{m}_2^{8,1} - \frac{3}{2N_c^2} \tilde{m}_3^{8,1} + \frac{1}{2N_c} \tilde{m}_2^{8,0} - \frac{1}{2N_c^2} \tilde{m}_1^{10 + \overline{10},1} + \frac{13}{20N_c} \tilde{m}_1^{27,2} \nonumber \\
& & \mbox{} + \frac{13}{20N_c^2} \tilde{m}_2^{27,2} - \frac{3}{20N_c} \tilde{m}_1^{27,0} - \frac{3}{20N_c^2} \tilde{m}_2^{27,0},
\end{eqnarray}
\begin{eqnarray}
M_{\Xi^-} & = & N_c \tilde{m}_1^{1,0} + \frac{3}{4N_c} \tilde{m}_2^{1,0} - \frac12 \tilde{m}_1^{8,1} + \frac{1}{4N_c} \tilde{m}_2^{8,1} - \frac{3}{4N_c^2} \tilde{m}_3^{8,1} - \frac12 \tilde{m}_1^{8,0} - \frac{3}{4N_c} \tilde{m}_2^{8,0} - \frac{3}{4N_c^2} \tilde{m}_3^{8,0} \nonumber \\
& & \mbox{} + \frac{1}{2N_c^2} \tilde{m}_1^{10 + \overline{10},1} + \frac{1}{20N_c} \tilde{m}_1^{27,2} + \frac{1}{20N_c^2} \tilde{m}_2^{27,2} + \frac{1}{5N_c} \tilde{m}_1^{27,1} + \frac{1}{5N_c^2} \tilde{m}_2^{27,1} + \frac{9}{20N_c} \tilde{m}_1^{27,0} \nonumber \\
& & \mbox{} + \frac{9}{20N_c^2} \tilde{m}_2^{27,0},
\end{eqnarray}
\begin{eqnarray}
M_{\Xi^0} & = & N_c \tilde{m}_1^{1,0} + \frac{3}{4N_c} \tilde{m}_2^{1,0} + \frac12 \tilde{m}_1^{8,1} - \frac{1}{4N_c} \tilde{m}_2^{8,1} + \frac{3}{4N_c^2} \tilde{m}_3^{8,1} - \frac12 \tilde{m}_1^{8,0} - \frac{3}{4N_c} \tilde{m}_2^{8,0} - \frac{3}{4N_c^2} \tilde{m}_3^{8,0} \nonumber \\
& & \mbox{} - \frac{1}{2N_c^2} \tilde{m}_1^{10 + \overline{10},1} + \frac{1}{20N_c} \tilde{m}_1^{27,2} + \frac{1}{20N_c^2} \tilde{m}_2^{27,2} - \frac{1}{5N_c} \tilde{m}_1^{27,1} - \frac{1}{5N_c^2} \tilde{m}_2^{27,1} + \frac{9}{20N_c} \tilde{m}_1^{27,0} \nonumber \\
& & \mbox{} + \frac{9}{20N_c^2} \tilde{m}_2^{27,0},
\end{eqnarray}
\begin{equation}
M_{\Lambda} = N_c \tilde{m}_1^{1,0} + \frac{3}{4N_c} \tilde{m}_2^{1,0} - \frac{1}{2N_c} \tilde{m}_2^{8,0} - \frac{3}{20N_c} \tilde{m}_1^{27,2} - \frac{3}{20N_c^2} \tilde{m}_2^{27,2} - \frac{27}{20N_c} \tilde{m}_1^{27,0} - \frac{27}{20N_c^2} \tilde{m}_2^{27,0},
\end{equation}
\begin{equation}
\sqrt{3} M_{\Sigma^0\Lambda} = \frac{3}{2N_c} \tilde{m}_2^{8,1} - \frac{3}{5N_c} \tilde{m}_1^{27,1} - \frac{3}{5N_c^2} \tilde{m}_2^{27,1} - \frac{3}{4N_c^2} \tilde{m}_2^{10 + \overline{10},3} + \frac{3}{4N_c^2} \tilde{m}_2^{10 + \overline{10},1},
\end{equation}
\begin{eqnarray}
M_{\Delta^{++}} & = & N_c \tilde{m}_1^{1,0} + \frac{15}{4N_c} \tilde{m}_2^{1,0} + \frac32 \tilde{m}_1^{8,1} + \frac{15}{4N_c} \tilde{m}_2^{8,1} + \frac{45}{4N_c^2} \tilde{m}_3^{8,1} + \frac12 \tilde{m}_1^{8,0} + \frac{5}{4N_c} \tilde{m}_2^{8,0} + \frac{15}{4N_c^2} \tilde{m}_3^{8,0} \nonumber \\
& & \mbox{} + \frac{21}{10N_c} \tilde{m}_1^{27,2} + \frac{21}{4N_c^2} \tilde{m}_2^{27,2} + \frac{3}{5N_c} \tilde{m}_1^{27,1} + \frac{3}{2N_c^2} \tilde{m}_2^{27,1} + \frac{9}{10N_c} \tilde{m}_1^{27,0} + \frac{9}{4N_c^2} \tilde{m}_2^{27,0} \nonumber \\
& & \mbox{} + \frac{9}{7N_c^2} \tilde{m}_1^{64,3} + \frac{11}{35N_c^2} \tilde{m}_1^{64,2} + \frac{3}{7N_c^2} \tilde{m}_1^{64,1} + \frac{9}{35N_c^2} \tilde{m}_1^{64,0},
\end{eqnarray}
\begin{eqnarray}
M_{\Delta^+} & = & N_c \tilde{m}_1^{1,0} + \frac{15}{4N_c} \tilde{m}_2^{1,0} + \frac12 \tilde{m}_1^{8,1} + \frac{5}{4N_c} \tilde{m}_2^{8,1} + \frac{15}{4N_c^2} \tilde{m}_3^{8,1} + \frac12 \tilde{m}_1^{8,0} + \frac{5}{4N_c} \tilde{m}_2^{8,0} + \frac{15}{4N_c^2} \tilde{m}_3^{8,0} \nonumber \\
& & \mbox{} - \frac{19}{10N_c} \tilde{m}_1^{27,2} - \frac{19}{4N_c^2} \tilde{m}_2^{27,2} + \frac{1}{5N_c} \tilde{m}_1^{27,1} + \frac{1}{2N_c^2} \tilde{m}_2^{27,1} + \frac{9}{10N_c} \tilde{m}_1^{27,0} + \frac{9}{4N_c^2} \tilde{m}_2^{27,0} \nonumber \\
& & \mbox{} - \frac{25}{7N_c^2} \tilde{m}_1^{64,3} - \frac{9}{35N_c^2} \tilde{m}_1^{64,2} + \frac{1}{7N_c^2} \tilde{m}_1^{64,1} + \frac{9}{35N_c^2} \tilde{m}_1^{64,0},
\end{eqnarray}
\begin{eqnarray}
M_{\Delta^0} & = & N_c \tilde{m}_1^{1,0} + \frac{15}{4N_c} \tilde{m}_2^{1,0} - \frac12 \tilde{m}_1^{8,1} - \frac{5}{4N_c} \tilde{m}_2^{8,1} - \frac{15}{4N_c^2} \tilde{m}_3^{8,1} + \frac12 \tilde{m}_1^{8,0} + \frac{5}{4N_c} \tilde{m}_2^{8,0} + \frac{15}{4N_c^2} \tilde{m}_3^{8,0} \nonumber \\
& & \mbox{} - \frac{19}{10N_c} \tilde{m}_1^{27,2} - \frac{19}{4N_c^2} \tilde{m}_2^{27,2} - \frac{1}{5N_c} \tilde{m}_1^{27,1} - \frac{1}{2N_c^2} \tilde{m}_2^{27,1} + \frac{9}{10N_c} \tilde{m}_1^{27,0} + \frac{9}{4N_c^2} \tilde{m}_2^{27,0} \nonumber \\
& & \mbox{} + \frac{25}{7N_c^2} \tilde{m}_1^{64,3} - \frac{9}{35N_c^2} \tilde{m}_1^{64,2} - \frac{1}{7N_c^2} \tilde{m}_1^{64,1} + \frac{9}{35N_c^2} \tilde{m}_1^{64,0},
\end{eqnarray}
\begin{eqnarray}
M_{\Delta^-} & = & N_c \tilde{m}_1^{1,0} + \frac{15}{4N_c} \tilde{m}_2^{1,0} - \frac32 \tilde{m}_1^{8,1} - \frac{15}{4N_c} \tilde{m}_2^{8,1} - \frac{45}{4N_c^2} \tilde{m}_3^{8,1} + \frac12 \tilde{m}_1^{8,0} + \frac{5}{4N_c} \tilde{m}_2^{8,0} + \frac{15}{4N_c^2} \tilde{m}_3^{8,0} \nonumber \\
& & \mbox{} + \frac{21}{10N_c} \tilde{m}_1^{27,2} + \frac{21}{4N_c^2} \tilde{m}_2^{27,2} - \frac{3}{5N_c} \tilde{m}_1^{27,1} - \frac{3}{2N_c^2} \tilde{m}_2^{27,1} + \frac{9}{10N_c} \tilde{m}_1^{27,0} + \frac{9}{4N_c^2} \tilde{m}_2^{27,0} \nonumber \\
& & \mbox{} - \frac{9}{7N_c^2} \tilde{m}_1^{64,3} + \frac{11}{35N_c^2} \tilde{m}_1^{64,2} - \frac{3}{7N_c^2} \tilde{m}_1^{64,1} + \frac{9}{35N_c^2} \tilde{m}_1^{64,0},
\end{eqnarray}
\begin{eqnarray}
M_{{\Sigma^*}^+} & = & N_c \tilde{m}_1^{1,0} + \frac{15}{4N_c} \tilde{m}_2^{1,0} + \tilde{m}_1^{8,1} + \frac{5}{2N_c} \tilde{m}_2^{8,1} + \frac{15}{2N_c^2} \tilde{m}_3^{8,1} + \frac{1}{2N_c} \tilde{m}_1^{27,2} + \frac{5}{4N_c^2} \tilde{m}_2^{27,2} - \frac{3}{5N_c} \tilde{m}_1^{27,1} \nonumber \\
& & \mbox{} - \frac{3}{2N_c^2} \tilde{m}_2^{27,1} - \frac{3}{2N_c} \tilde{m}_1^{27,0} - \frac{15}{4N_c^2} \tilde{m}_2^{27,0} - \frac{2}{7N_c^2} \tilde{m}_1^{64,3} - \frac{24}{35N_c^2} \tilde{m}_1^{64,2} - \frac{10}{7N_c^2} \tilde{m}_1^{64,1} \nonumber \\
& & \mbox{} - \frac{36}{35N_c^2} \tilde{m}_1^{64,0},
\end{eqnarray}
\begin{eqnarray}
M_{{\Sigma^*}^0} & = & N_c \tilde{m}_1^{1,0} + \frac{15}{4N_c} \tilde{m}_2^{1,0} - \frac{3}{2N_c} \tilde{m}_1^{27,2} - \frac{15}{4N_c^2} \tilde{m}_2^{27,2} - \frac{3}{2N_c} \tilde{m}_1^{27,0} - \frac{15}{4N_c^2} \tilde{m}_2^{27,0} + \frac{36}{35N_c^2} \tilde{m}_1^{64,2} \nonumber \\
& & \mbox{} - \frac{36}{35N_c^2} \tilde{m}_1^{64,0},
\end{eqnarray}
\begin{eqnarray}
M_{{\Sigma^*}^-} & = & N_c \tilde{m}_1^{1,0} + \frac{15}{4N_c} \tilde{m}_2^{1,0} - \tilde{m}_1^{8,1} - \frac{5}{2N_c} \tilde{m}_2^{8,1} - \frac{15}{2N_c^2} \tilde{m}_3^{8,1} + \frac{1}{2N_c} \tilde{m}_1^{27,2} + \frac{5}{4N_c^2} \tilde{m}_2^{27,2} + \frac{3}{5N_c} \tilde{m}_1^{27,1} \nonumber \\
& & \mbox{} + \frac{3}{2N_c^2} \tilde{m}_2^{27,1} - \frac{3}{2N_c} \tilde{m}_1^{27,0} - \frac{15}{4N_c^2} \tilde{m}_2^{27,0} + \frac{2}{7N_c^2} \tilde{m}_1^{64,3} - \frac{24}{35N_c^2} \tilde{m}_1^{64,2} + \frac{10}{7N_c^2} \tilde{m}_1^{64,1} \nonumber \\
& & \mbox{} - \frac{36}{35N_c^2} \tilde{m}_1^{64,0},
\end{eqnarray}
\begin{eqnarray}
M_{{\Xi^*}^-} & = & N_c \tilde{m}_1^{1,0} + \frac{15}{4N_c} \tilde{m}_2^{1,0} - \frac12 \tilde{m}_1^{8,1} - \frac{5}{4N_c} \tilde{m}_2^{8,1} - \frac{15}{4N_c^2} \tilde{m}_3^{8,1} - \frac12 \tilde{m}_1^{8,0} - \frac{5}{4N_c} \tilde{m}_2^{8,0} - \frac{15}{4N_c^2} \tilde{m}_3^{8,0} \nonumber \\
& & \mbox{} - \frac{1}{10N_c} \tilde{m}_1^{27,2} - \frac{1}{4N_c^2} \tilde{m}_2^{27,2} + \frac{4}{5N_c} \tilde{m}_1^{27,1} + \frac{2}{N_c^2} \tilde{m}_2^{27,1} - \frac{9}{10N_c} \tilde{m}_1^{27,0} - \frac{9}{4N_c^2} \tilde{m}_2^{27,0} \nonumber \\
& & \mbox{} - \frac{2}{7N_c^2} \tilde{m}_1^{64,3} + \frac{6}{35N_c^2} \tilde{m}_1^{64,2} - \frac{10}{7N_c^2} \tilde{m}_1^{64,1} + \frac{54}{35N_c^2} \tilde{m}_1^{64,0},
\end{eqnarray}
\begin{eqnarray}
M_{{\Xi^*}^0} & = & N_c \tilde{m}_1^{1,0} + \frac{15}{4N_c} \tilde{m}_2^{1,0} + \frac12 \tilde{m}_1^{8,1} + \frac{5}{4N_c} \tilde{m}_2^{8,1} + \frac{15}{4N_c^2} \tilde{m}_3^{8,1} - \frac12 \tilde{m}_1^{8,0} - \frac{5}{4N_c} \tilde{m}_2^{8,0} - \frac{15}{4N_c^2} \tilde{m}_3^{8,0} \nonumber \\
& & \mbox{} - \frac{1}{10N_c} \tilde{m}_1^{27,2} - \frac{1}{4N_c^2} \tilde{m}_2^{27,2} - \frac{4}{5N_c} \tilde{m}_1^{27,1} - \frac{2}{N_c^2} \tilde{m}_2^{27,1} - \frac{9}{10N_c} \tilde{m}_1^{27,0} - \frac{9}{4N_c^2} \tilde{m}_2^{27,0} \nonumber \\
& & \mbox{} + \frac{2}{7N_c^2} \tilde{m}_1^{64,3} + \frac{6}{35N_c^2} \tilde{m}_1^{64,2} + \frac{10}{7N_c^2} \tilde{m}_1^{64,1} + \frac{54}{35N_c^2} \tilde{m}_1^{64,0},
\end{eqnarray}
\begin{eqnarray}
M_{\Omega^-} & = & N_c \tilde{m}_1^{1,0} + \frac{15}{4N_c} \tilde{m}_2^{1,0} - \tilde{m}_1^{8,0} - \frac{5}{2N_c} \tilde{m}_2^{8,0} - \frac{15}{2N_c^2} \tilde{m}_3^{8,0} + \frac{3}{10N_c} \tilde{m}_1^{27,2} + \frac{3}{4N_c^2} \tilde{m}_2^{27,2} + \frac{27}{10N_c} \tilde{m}_1^{27,0} \nonumber \\
& & \mbox{} + \frac{27}{4N_c^2} \tilde{m}_2^{27,0} - \frac{4}{35N_c^2} \tilde{m}_1^{64,2} - \frac{36}{35N_c^2} \tilde{m}_1^{64,0}.
\end{eqnarray}

\end{document}